\documentclass[journal]{IEEEtran}
\usepackage{color,soul}
\usepackage{psfrag}
\usepackage{epsfig}
\usepackage{breakcites}
\usepackage{graphicx}
\usepackage[caption=false]{subfig}
\usepackage{bm}
\usepackage{cite}
\usepackage{amsmath,amssymb,amsfonts,amsthm}
\usepackage{balance}
\usepackage{textcomp}
\usepackage{cite} 
\usepackage{float,algorithm,algpseudocode}
\usepackage{amsopn} 
\usepackage{tikz}

\graphicspath{{./img/}}
\DeclareGraphicsExtensions{.pdf,.jpeg,.png}
\usepackage{mathtools} 
\usepackage{xspace}

\newcommand{\Figref}[1]{Fig.~\ref{#1}}

\newcommand{\ba}{\mathbf{a}}

\newcommand{\bg}{\mathbf{g}}
\newcommand{\bh}{\mathbf{h}}

\newcommand{\bv}{\mathbf{v}}

\newcommand{\bz}{\mathbf{z}}

\newcommand{\bA}{\mathbf{A}}
\newcommand{\bB}{\mathbf{B}}
\newcommand{\bC}{\mathbf{C}}
\newcommand{\bD}{\mathbf{D}}

\newcommand{\bF}{\mathbf{F}}
\newcommand{\bG}{\mathbf{G}}

\newcommand{\bI}{\mathbf{I}}
\newcommand{\bJ}{\mathbf{J}}

\newcommand{\bP}{\mathbf{P}}

\newcommand{\bR}{\mathbf{R}}

\newcommand{\bzero}{\boldsymbol{0}}

\newtheorem{proposition}{Proposition}

\newtheorem{remark}{Remark}

\newcommand{\diag}{\text{diag}}

\newcommand{\herm}{^{\mathsf{H}}}
\newcommand{\trans}{^\mathsf{T}}

\DeclareMathOperator{\tr}{tr}
\DeclareMathOperator{\E}{\mathsf{E}}

\newcommand{\EX}[1]{\E\left\{{#1}\right\}}

\DeclareMathOperator*{\argmax}{arg\,max}

\newcommand{\euler}{\mathrm{e}}

\newcommand{\norm}[1]{{ \left\Vert #1 \right\Vert }}

\newcommand{\mC}{\mathbb{C}}

\newcommand{\CG}[2]{\mathcal{CN}\left({#1},{#2}\right)}

\newcommand{\dFA}{d_{\text{FA}}}
\newcommand{\dH}{d_{\text{H}}^{\texttt{i}_{n,q},\texttt{i}_{n',q'}}}
\newcommand{\dV}{d_{\text{V}}^{\texttt{i}_{n,q},\texttt{i}_{n',q'}}}



\begin{document}
	\IEEEoverridecommandlockouts
	
	\pagenumbering{gobble}
	
	This work has been submitted to the IEEE for possible publication. Copyright may be transferred without notice, after which this version may no longer be accessible.
	\newpage

\title{User-Centric Cell-Free Massive MIMO Enhanced by Fluid-Antenna Access Points: Uplink Analysis}

\author{Maryam Olyaee, Giovanni Interdonato, \IEEEmembership{Member,~IEEE}, and Stefano Buzzi,~\IEEEmembership{Fellow,~IEEE}\vspace{-5mm}

\thanks{A preliminary and shortened version of the technical content presented in this paper is reported in~\cite{olyaee2024user}.}
\thanks{M. Olyaee was supported by the Horizon Europe Programme Marie Skłodowska-Curie Actions Postdoctoral Fellowships (Grant No. 101107131).
S. Buzzi and G. Interdonato were supported by the European Union under the Italian National Recovery and Resilience Plan (NRRP) of NextGenerationEU, partnership on ``Telecommunications of the Future'' (PE00000001 - program ``RESTART'', Structural Project 6GWINET, Cascade Call SPARKS, CUP: D43C22003080001).}
\thanks{The authors are with the Dept. of Electrical and Information Engineering (DIEI), University of Cassino and Southern Lazio, 03043 Cassino, Italy (e-mail:\{maryam.olyaee; buzzi; giovanni.interdonato\}@unicas.it). S. Buzzi and G. Interdonato are also with the Consorzio Nazionale Interuniversitario per le Telecomunicazioni (CNIT), 43124 Parma, Italy. S. Buzzi is also with the Dipartimento di Elettronica, Informazione e Bioingegneria (DEIB), Politecnico di Milano, 20156 Milan, Italy.}
}

\maketitle

\begin{abstract}
In this paper, we investigate cell-free massive MIMO (CF-mMIMO) systems in which access points (APs) are equipped with fluid antennas (FAs) and develop a comprehensive framework for channel estimation, antenna port selection, and uplink spectral efficiency (SE) optimization.
We propose a generalized LMMSE-based uplink channel estimation scheme that dynamically activates FA ports during pilot transmission, efficiently exploiting antenna reconfigurability under practical training constraints. Building on this, we design a distributed port selection strategy that minimizes per-AP channel estimation error by exploiting spatial correlation among FA ports.
We systematically analyze the impact of antenna geometry and spatial correlation using the Jakes’ channel model for different AP array configurations, including uniform linear and planar arrays. We then derive SINR expressions for centralized and distributed uplink processing and obtain a closed-form uplink SE expression for centralized maximum-ratio combining using the use-and-then-forget bound. Finally, we propose an alternating-optimization framework to select FA port configurations that maximize the uplink sum SE.
Numerical results show that the proposed FA-aware channel estimation and port optimization strategies greatly reduce channel estimation error and significantly improve sum-SE over fixed-antenna and non-optimized FA baselines, confirming FAs as a key enabler for scalable, adaptive CF-mMIMO networks.
\end{abstract}
	
\section{Introduction}

In recent years, a novel antenna paradigm known as the \emph{fluid antenna} (FA), also referred to as a movable antenna, has emerged as a promising solution for enhancing the performance and flexibility of wireless communication systems~\cite{zhu2023movable,wong2020fluid}. Unlike conventional antennas with fixed physical locations, FAs are realized using fluidic, conductive, and dielectric materials whose effective geometry and radiating position can be dynamically reconfigured through software control~\cite{wong2023fluid}. A representative practical implementation consists of a droplet of liquid metal, such as eutectic gallium--indium (EGaIn), confined within a microfluidic structure filled with an electrolyte. By exploiting electromagnetic, electrocapillary, or pressure-based actuation mechanisms, the liquid metal droplet can be repositioned within the structure, thereby modifying the antenna radiation point~\cite{fe6}.

This reconfigurability is particularly appealing in wireless environments, since the propagation channel exhibits significant spatial variations over distances on the order of the signal wavelength. As a result, even small displacements of the antenna can induce substantial changes in the channel response, which can be exploited to improve signal quality, mitigate deep fades, and enhance spatial diversity. Leveraging this property, FA systems have recently been investigated for a wide range of wireless communication scenarios, including point-to-point links, conventional multiple-input multiple-output (MIMO) systems, and, more recently, user-centric cell-free massive MIMO (CF-mMIMO) architectures~\cite{wong2020fluid,M1}.

The integration of FAs into user-centric CF-mMIMO networks~\cite{CFNgo2016,buzzi2017cell} is particularly attractive, as it enables spatial adaptability at the infrastructure level while preserving the distributed nature of the system. In CF-mMIMO, a large number of geographically distributed access points (APs), each equipped with a limited number of antennas, jointly and coherently serve the users without cell boundaries. Compared to traditional multi-cell massive MIMO systems with centralized antenna arrays, this architecture offers more uniform quality of service, improved macro-diversity, and enhanced robustness to shadowing and interference. Owing to these properties, CF-mMIMO is widely regarded as a key enabling technology for future 6G wireless networks~\cite{demir2021foundations,ngo2024ultradense,BuzziCM2026}.

In this context, FAs add a degree of freedom by enabling dynamic adaptation of each AP’s antenna radiation point to the propagation environment and user distribution. Initial work in this direction appeared in~\cite{olyaee2024user}, which studied uplink power control and antenna position optimization in FA-aided CF-mMIMO systems. This work substantially extends that study by proposing a comprehensive framework for channel estimation, antenna port selection, and spectral efficiency optimization in CF-mMIMO with FAs.

\subsection{Related Works and Motivation}
Recent years have witnessed growing interest in the use of FAs as a means to exploit spatial channel variations for interference mitigation and massive connectivity. A major line of research has focused on \emph{fluid antenna multiple access} (FAMA), where users are equipped with a single FA and dynamically select the antenna port that offers favorable interference conditions.

In this context, the slow fluid antenna multiple access (s-FAMA) paradigm has been introduced and analyzed in several works~\cite{wong2023slow,vega2025tractable,yang2023performance}. In s-FAMA, the active FA port is updated only when the large-scale fading conditions change, significantly reducing the port-switching overhead. These studies provide analytical characterizations of outage probability and multiplexing gains, demonstrating that s-FAMA can support high spatial multiplexing, provided that a sufficiently large number of FA ports is available. 

Beyond FAMA, FAs have also been investigated in conventional multiple access settings. For example, \cite{new2023fluid} studies FA-assisted orthogonal and non-orthogonal multiple access (OMA/NOMA) systems, formulating joint port selection and power allocation problems to maximize the sum rate. The results show that FA-enabled systems can significantly outperform traditional antenna selection schemes, further highlighting the potential of FA technology for massive connectivity in future wireless networks.

More recently, efforts have started to explore the integration of FAs into distributed and cell-free architectures. In particular, \cite{han2025cell} extends the FAMA concept to a cell-free setting by equipping users with a single FA while access points (APs) employ fixed antennas and maximum-ratio transmission. The analysis mainly focuses on outage probability under different port-switching strategies, while crucial aspects such as channel estimation and infrastructure-side antenna reconfigurability are \textit{not} addressed. As a result, the potential of FAs at the AP level within user-centric CF-mMIMO systems remains largely unexplored.

A fundamental challenge in realizing FA-enabled distributed MIMO networks lies in the design of optimization strategies that can efficiently exploit spatial adaptability while ensuring scalability and fairness across users. A key enabler of such optimization is the availability of accurate CSI at the antenna ports. In our prior work~\cite{olyaee2024user}, uplink power control and antenna position optimization were studied under the assumption of perfect CSI. While this assumption facilitates initial insights, it overlooks one of the most critical practical bottlenecks in FA systems, namely CSI acquisition.

Indeed, channel estimation in FA systems is inherently challenging. On the one hand, the potentially large number of antenna ports leads to extremely high-dimensional channel vectors. On the other hand, due to the limited number of RF chains, only a small subset of ports can be observed at any given time, rendering exhaustive port-by-port estimation impractical. Consequently, efficient CSI acquisition strategies that balance estimation accuracy and training overhead are essential.

Several recent works have addressed CSI estimation for FA systems. In \cite{xu2023channel}, sparse channel reconstruction techniques are proposed for mmWave systems with FA-equipped users. Compressed sensing-based approaches are investigated in~\cite{ma2023compressed} to jointly estimate the channel and optimize antenna positioning while reducing pilot overhead. Other studies exploit structural properties of FA channels: regression-based methods are proposed in~\cite{wang2023estimation} to reduce the number of required channel measurements, while machine learning techniques are employed in~\cite{ji2024correlation} to reconstruct full channel responses from limited port observations. Sparse Bayesian learning approaches are developed in~\cite{ML2}, showing notable improvements over conventional estimation schemes.

Despite these advances, a common limitation of existing FA channel estimation methods is the significant pilot overhead and computational complexity incurred when the number of antenna ports is large. To mitigate this issue, \cite{fe6} proposes MMSE-based sequential estimation strategies that selectively probe a subset of antenna ports, leveraging spatial correlation to infer the remaining channels. While promising, such approaches have so far been studied primarily in centralized or single-cell scenarios.

Despite the substantial body of work on FA systems, several important gaps remain. 
First, most existing studies consider FA-enabled multiple access schemes in centralized or single-cell settings, with FAs located at the user side, while the exploitation of FAs at the infrastructure level in distributed, user-centric CF-mMIMO networks has received very limited attention. 
Second, the majority of available results either neglect channel estimation altogether or assume idealized CSI acquisition, which is particularly unrealistic in FA systems due to the large number of potential antenna ports and the limited number of RF chains. 
Third, existing FA-related optimization frameworks typically focus on isolated aspects such as outage probability, port selection, or power allocation, without jointly accounting for channel estimation accuracy, antenna spatial correlation, and scalable signal processing in large networks. 
As a result, a unified and practically viable framework that integrates FA-aware channel estimation, antenna port selection, and uplink performance optimization in user-centric CF-mMIMO systems is still missing.

\subsection{Main Contributions}

This work investigates the fundamental problem of channel estimation and antenna reconfiguration in user-centric CF-mMIMO systems equipped with FAs at the distributed APs. The presence of FAs introduces new degrees of freedom through port reconfigurability, but also raises nontrivial challenges related to channel estimation overhead, spatial correlation, and scalable signal processing. The main contributions of this paper can be summarized as follows.

\begin{itemize}

\item[-] We introduce and study a user-centric CF-mMIMO architecture in which multi-antenna APs are equipped with FAs. This represents, to the best of our knowledge, the first comprehensive investigation of FA technology within a CF-mMIMO framework, shifting spatial adaptability from the user side to the network infrastructure and enabling scalable, reconfigurable distributed MIMO operation.

\item[-] We propose a generalized uplink channel estimation framework based on linear minimum mean-squared error (LMMSE) estimation tailored to FA-equipped APs. The framework allows the active antenna port to vary across pilot symbols, enabling sequential probing of multiple FA ports and, in principle, full CSI acquisition at the cost of increased training overhead. As a low-complexity baseline, we also consider fixed-port training, which significantly reduces overhead while still capturing the essential behavior of FA-based systems.

\item[-] To efficiently exploit the available FA ports, we develop a port selection strategy that minimizes the per-AP normalized mean-squared error (NMSE) of the channel estimates. The resulting algorithm, referred to as \emph{local NMSE-descent port selection} (LND-PS), enables distributed and scalable optimization of FA port configurations based solely on local channel statistics, making it well suited for large CF-mMIMO deployments.

\item[-] We explicitly account for antenna spatial correlation and investigate its impact on channel estimation and achievable rate performance. Using the Jakes’ correlation model, we analyze how FA port spacing and array geometry affect correlation properties. In particular, we study uniform linear (ULA), uniform planar (UPA), and uniform circular (UCA) array configurations, highlighting the critical role of array geometry in determining estimation accuracy, spatial diversity, and the effectiveness of FA port optimization.

\item[-] We derive signal-to-interference-plus-noise ratio (SINR) expressions for both centralized and distributed uplink processing. In addition, we obtain a closed-form SINR expression for maximum-ratio combining (MRC) using the widely adopted \textit{Use-and-Then-Forget} (UatF) capacity-bounding technique. These analytical results provide useful performance benchmarks and offer insight into the interplay between estimation quality, antenna correlation, and FA reconfiguration.

\item[-] Building on the derived SINR expressions, we formulate and solve an optimization problem for FA port selection aimed at maximizing the uplink sum spectral efficiency (SE). An alternate optimization (AO) algorithm is proposed to jointly account for channel estimation quality and data reception, demonstrating how dynamic FA port selection can be effectively leveraged to enhance network throughput.

\item[-] Finally, we present an extensive simulation campaign that validates the proposed analytical framework and optimization strategies. The results demonstrate significant gains in channel estimation accuracy, sum SE, and user fairness compared to conventional fixed-antenna and non-optimized FA baselines, thereby confirming the potential of FAs as an enabling technology for next-generation CF-mMIMO systems.

\end{itemize}

It is finally worth emphasizing that, although this paper focuses on FAs, the proposed framework and methodologies readily extend to other emerging reconfigurable antenna technologies, including pinchable antenna arrays \cite{tyrovolas2025performance} and meta–FA architectures \cite{liu2025meta}.

\subsection{Paper Organization and Notation}
The rest of the paper is organized as follows. Section \ref{sys} introduces the system model, while Section \ref{estimation_sec} presents the proposed channel estimation method.
Section \ref{metric} focuses on the uplink signal processing, introduces the performance metrics of our interest, and describes the proposed AO algorithm for antenna port selection.
Numerical results are discussed in Section \ref{numerical}. Finally, Section \ref{concl} concludes the paper.

\textit{Notation:} Lower-case boldface letters, $\ba$, denote vectors, while upper-case boldface letters, $\bA$, are used for matrices. The transpose, inverse, and conjugate transpose of a matrix $\bA$ are represented by $\bA\trans\!$, $\bA^{-1}$, and $\bA\herm$, respectively.
The $n$-th column and $n$-th row of $\bA$ are denoted by $[\bA]_{:,n}$ and $[\bA]_{n,:}$, respectively, and the $n$-th element of a vector $\ba$ is written as $[\ba]_n$. A diagonal matrix formed by the scalars $a_1, \ldots, a_N$ is expressed as $\diag(a_1, \ldots, a_N)$. The trace of a matrix $\bA$ is denoted by $\tr\{\bA\}$.
The sets of $N$ different instances of the vector $\ba$ and of the matrix $\bA$ are denoted by $\{\ba_n\}_{n=1}^N$ and $\{\bA(n)\}_{n=1}^N$, respectively. The $N$-dimensional identity matrix is denoted by $\bI_N$. A complex circularly symmetric Gaussian random variable with mean $\mu$ and variance $\sigma^2$ is expressed as $\mathcal{CN} \sim (\mu, \sigma^2)$. The symbol $\mathbb{C}
$ represents the set of complex numbers.
 The statistical expectation operator is denoted by $\EX{\cdot}$, the Kronecker product is represented by $\otimes$, and the number of nonzero elements in $\ba$ represented by 
$\|\ba\|_0$. Lastly, we denote the covariance operator by $\text{cov}\{\cdot\}$ and the cardinality of a set $\mathcal{A}$ by $|\mathcal{A}|$.

\section{System Model}\label{sys}
Consider a CF-mMIMO system with $L$ APs, each one equipped with $N$ FAs, serving $K$ users in the same time-frequency resource. We assume that $NL > K$.
The APs are assumed to be connected through fronthaul links to a central processing unit (CPU).
We follow the user-centric approach \cite{buzzi2017cell}, i.e., we assume that each user is served by a limited number of APs, which are selected according to the largest values of the large-scale fading coefficients. Accordingly, we denote by $\mathcal{L}_k$  the set of APs that communicate with the $k$-th user, and by $\mathcal{K}_l$  the set of users served by the $l$-th AP. Thus the set 
$\mathcal{L}_k$ can be defined as 
$\mathcal{L}_k=\{l:k \in \mathcal{K}_l \}$; moreover, the association among APs and users can also be encoded in a binary $(K \times L)$-dimensional matrix, whose $(k,l)$-th element is 1 if $k \in \mathcal{K}_l$ and 0 otherwise. The cardinality of the set $\mathcal{L}_k$, $|\mathcal{L}_k|$, represents the number of APs serving user $k$.

We assume an FA architecture consisting of a drop of liquid metal (that is, the FA itself) placed in a tube-like linear microchannel or capillary filled with an electrolyte, within which the fluid is free to move. Each FA is located in one of the preset $Q$ locations, also referred to as the FA \textit{ports}, that are evenly distributed along the linear dimension of an FA. As already commented, our approach straightforwardly extend to other movable antenna architectures, such as pinchable antennas \cite{tyrovolas2025performance}.

\subsection{Channel Model}
Let $h_{k,l,n,q} \!\in \mC$ be the complex channel coefficient between user $k$ and AP $l$ through the $q$-th port of the $n$-th FA, and $\bh_{k,l,n} =\! [h_{k,l,n,1}, \ldots, h_{k,l,n,Q}]\trans\!\in \mC^{Q}$ be the auxiliary channel vector between user $k$ and AP $l$ through the $n$-th FA, which is obtained by stacking the coefficients $\{h_{k,l,n,q}\}^Q_{q=1}$. 
%
By stacking the vectors $\bh_{k,l,n}$, $\forall n$, the resulting collective auxiliary vector of the channel between user $k$ and AP $l$ is given by $\bh_{k,l} \!=\! [\bh\trans_{k,l,1}, \ldots, \bh\trans_{k,l,N}]\trans \!\in \mC^{NQ}$.
%
Let $\texttt{i}_{n,q}\!=\!\{1,\ldots,NQ\}$ be the unique index assigned to the $q$-th port of FA $n$ in any AP, given by
\begin{equation}\label{inq}
    \texttt{i}_{n,q} \!=\! (n\!-\!1)Q\!+\!q\, ,
\end{equation}
then, we define the mapping $h_{k,l,n,q} \longleftrightarrow \dot{h}_{k,l,{\texttt{i}_{n,q}}}$ and 
\begin{align}
\bh_{k,l} \!=\! [\dot{h}_{k,l,{\texttt{i}_{1,1}}},\ldots,\dot{h}_{k,l,{\texttt{i}_{N,Q}}}]\trans \!\in\! \mC^{NQ}\,.
\end{align}
As the ports within an AP are closely located, the entries of $\bh_{k,l}$ are correlated. 
Unlike the vast majority of prior works on FA systems that have primarily considered LoS channels, we assume NLoS channels, targeting urban CF-mMIMO deployments with dense scattering and pronounced shadowing, thereby leading to weak or blocked LoS components.
We capture the spatial correlation between different ports over different FAs by modeling the collective auxiliary channel vector as $\bh_{k,l} \sim \CG{\bzero_{NQ}}{\bR_{k,l}}$,
where $\bR_{k,l}\!=\!\{\bh_{k,l}\bh\herm_{k,l}\}\!\in\!\mC^{NQ\times NQ}$ is the auxiliary block-diagonal spatial correlation matrix.
While~\cite{olyaee2024user} employed the local scattering model together with Gaussian angular distributions for the multipath components~\cite{demir2021foundations,Emil1}, this paper refines the modeling of small-scale fading characteristics, particularly the spatial correlation between ports within the same AP, by adopting the Jakes' model due to its capability to capture isotropic propagation effects~\cite{tot,jake}.
Despite its significance and widespread adoption, the Jakes' model, although it accounts for array geometry, does not consider the angular distribution of channel multipath components. To achieve a more comprehensive analysis, we therefore evaluate system performance under both the Jakes' and local scattering models, drawing insights from their comparative assessment.

Generalizing, the $({\texttt{i}_{n,q},\texttt{i}_{n',q'}})$-th element of the spatial correlation matrix $\mathbf{R}_{k,l}$ is given by
\begin{align}
    [\mathbf{R}_{k,l}]_{\texttt{i}_{n,q},\texttt{i}_{n',q'}} &=  \text{cov}\{ \dot{h}_{k,l,{\texttt{i}_{n,q}}},\dot{h}_{k,l,{\texttt{i}_{n',q'}}}\} \nonumber \\
    & = \beta_{k,l} \cdot[\mathbf{J}]_{\texttt{i}_{n,q},\texttt{i}_{n',q'}}\,,
    \label{RJ}
\end{align}
where $\bJ$ denotes the normalized spatial correlation matrix that accounts for the antenna geometry, while $\beta_{k,l}$ is the large-scale fading coefficient characterizing the channel from the $k$-th user to the $l$-th AP, which encompasses pathloss and shadowing, and is given by $\beta_{k,l}=\tr\{\mathbf{R}_{k,l}\}/NQ$. Let ${\bP}_{\text{x},l}$ denote the port position matrix, a binary $N \times NQ$-dimensional matrix encoding the position of the active ports in each of the $N$ FAs at the $l$-th AP. Specifically, the entry of ${\bP}_{\text{x},l}$ at row $n$ and column $(n-1)Q + q$ is 1 if the $q$-th port of the $n$-th antenna is active, and 0 otherwise.
The subscript $\text{x} \in \{\text{p},\text{d}\}$ specifies whether the FA ports are active during the uplink training, namely the antenna ports receive pilot sequences, or during the uplink data transmission phase, in which the antenna ports receive data. Accordingly, it holds that $\beta_{k,l}=\tr\{{\bP}_{\text{x},l}\mathbf{R}_{k,l}\}/N$.

\subsection{AP Geometries and Antenna-Port Spatial Correlation}
  
In this section, we examine the spatial correlation characteristics at the antenna port induced by the geometry of the access point.
We assume that every AP antenna is configured as a linear, cylinder-shaped FA with length $\lambda\ell$, where $\lambda$ is the wavelength and $\ell$ denotes the normalized FA length.
We assume a minimum distance between adjacent antennas equal to $\lambda/2$, so as to avoid mutual coupling effects. The FA port (i.e., the liquid metal within the antenna) can move over $\lambda\Delta Q$,  and has a length equal to $\lambda \Delta$, where $\Delta$ is also the normalized antenna spacing. Next, we characterize the spatial correlation between ports and antennas induced by the following three AP geometries: $(i)$ ULA, $(ii)$ UPA, and $(iii)$ UCA (also known as \textit{circular} FA~\cite{fe6}).

\subsubsection{ULA with FAs}
\begin{figure}[!t]
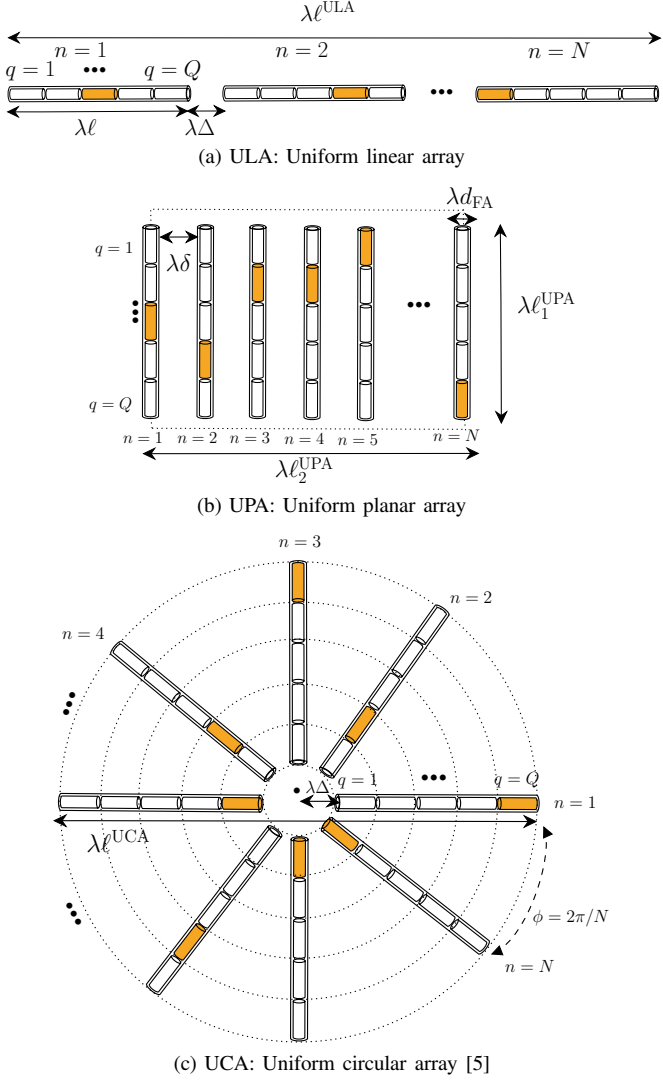
  
    \centering
    \subfloat[ULA: Uniform linear array]{\resizebox{\columnwidth}{!}{\input{ULA}}\label{geometry0}}%
    \vspace{2mm}
    \subfloat[UPA: Uniform planar array]{\resizebox{.75\columnwidth}{!}{\input{UPA}}\label{geometry1}}%
    \vspace{2mm}
    \subfloat[UCA: Uniform circular array \cite{fe6}]{\resizebox{.85\columnwidth}{!}{\input{UCA}}\label{circ}}%
    \caption{Geometries of FA-equipped AP: (a) ULA, (b) UPA, and (c) UCA.}
    \label{fig:FA_geometries}
\end{figure}

Let us assume that each AP is equipped with an $N$-antenna ULA. Then, the normalized length of the ULA is given by $\ell^{\text{ULA}} = N\ell+ (N-1)\Delta$, where $\ell = Q\Delta$, as illustrated in~\Figref{geometry0}. The $({\texttt{i}_{n,q},\texttt{i}_{n',q'}})$-th entry of the normalized spatial correlation matrix is, according to the Jakes' model and for any user-to-AP link, given by
\begin{align}
    [\mathbf{J}^{\text{ULA}}]_{\texttt{i}_{n,q},\texttt{i}_{n',q'}} & = 
    J_0\left( \frac{2 \pi \lvert\bar{\texttt{i}}_{n-n',q-q'}\rvert \ell^{\text{ULA}}}{QN+N - 2}  \right) \nonumber \\ 
    &= J_0\left( \frac{2 \pi  \lvert \overline{n}Q +\overline{p} \rvert \ell^{\text{ULA}}}{QN +N- 2}  \right),
    \label{eq:J:ULA:Jakes}
\end{align}
where $\bar{\texttt{i}}_{n-n',q-q'} = \texttt{i}_{n,q} - \texttt{i}_{n',q'}$, $\overline{n} = n-n'$, $\overline{p} = p-p'$, and $J_0(\cdot)$ is the zero-order Bessel function of the first kind.

Alternatively, the normalized spatial correlation matrix can be modeled according to the local scattering model~\cite{demir2021foundations}.
For a ULA deployed along a horizontal line with half-wavelength antenna spacing, and under the assumption that all multipath components arrive from the far-field, the normalized spatial correlation matrix is in~\eqref{eq:J:ULA:Jakes} is redefined as~\cite[eq. (2.18)]{demir2021foundations}
\begin{align}
    [\mathbf{J}^{\text{ULA}}]&_{\texttt{i}_{n,q},\texttt{i}_{n',q'}} \!=\! \int\!\!\! \int {\euler}^{j\pi \bar{\texttt{i}}_{n-n',q-q'}\sin(\bar\varphi)\cos(\bar\theta)}f(\bar\varphi,\bar\theta)\,d\bar\varphi\, d\bar\theta\,,
\end{align}
where $\bar\varphi$ denotes the azimuth angle and $\bar\theta$ denotes the elevation angle of a multipath component, and $f(\bar\varphi,\bar\theta)$ is the joint probability density
function (PDF) of $\bar\varphi$ and  $\bar\theta$.

\subsubsection{UPA with FAs}
Let us assume that each AP is equipped with an $N$-antenna UPA, as illustrated in~\Figref{geometry1}, whereby the FAs are arranged side by side and perpendicular to the ground. The vertical and horizontal apertures of the UPA are $\ell_1^{\text{UPA}} =\ell$ and $\ell_2^{\text{UPA}} = (N-1)\delta + N \dFA$, respectively, where $\delta$ is the normalized spacing between adjacent FAs and $\dFA$ is the normalized FA cross-section diameter, with $\dFA = \Delta - \delta$.
According to Jake's model, the $({\texttt{i}_{n,q},\texttt{i}_{n',q'}})$-th entry of the normalized spatial correlation matrix is, for any user-to-AP link, given by
\begin{align}
    [\mathbf{J}^{\text{UPA}}]_{\texttt{i}_{n,q} ,\texttt{i}_{n',q'} } \!=\!
    j_0\!\left(2\pi \sqrt{{\left(\frac{|\overline{q}|\ell_1^{\text{UPA}}}{Q\!-\!1}  \right)}^{\!\!2} \!+\! \left(\frac{|\overline{n}|\ell_2^{\text{UPA}}}{N\!-\!1}  \right)^{\!\!2}}\right),
\end{align}
where $j_0(\cdot)$ is the zero-order spherical Bessel function of the first kind.

Alternatively, the normalized spatial correlation matrix, in the case of UPAs, can be modeled according to the local scattering model as~\cite[Sec. 7.3]{Emil1},
\begin{align}
    \mathbf{J}^{\text{UPA}} = \int \int f(\bar\varphi,\bar\theta)\,\ba(\bar\varphi,\bar\theta)\,\ba\herm(\bar\varphi,\bar\theta)\,d\bar\varphi\,d\bar\theta\,,
    \label{eq:J:UPA:Jakes}
\end{align}
where $\ba(\bar\varphi,\bar\theta)$ is the array response vector. Assuming that the UPA is deployed along the $y$- and $z$-axes in the horizontal and vertical directions, respectively, the entries of $\ba(\bar\varphi,\bar\theta)$ depend on the indices $\texttt{i}_{n,q}$ and $\texttt{i}_{n',q'}$ and are given by
\begin{align}
    &[\ba(\bar\varphi,\bar\theta)]_{\texttt{i}_{n,q},\texttt{i}_{n',q'}} \nonumber \\
    &\quad = {\euler}^{j2\pi \dH \sin(\bar\varphi)\cos(\bar\theta)}  \,{\euler}^{j 2\pi\dV \sin(\bar\theta)}\,,
\end{align}
where $\dH$ and $\dV$ denote the horizontal and vertical distances, respectively, between the $\texttt{i}_{n,q}$-th and the $\texttt{i}_{n',q'}$-th antenna ports, expressed in wavelengths.

\subsubsection{UCA with FAs} 
Let us consider that each AP is equipped with an $N$-antenna UCA, as illustrated in~\Figref{circ}. The port $q$ at the $n$-th FA is located at polar coordinates $(q+\frac{1}{2})\Delta$ (radius) and $\frac{2\pi}{N}(n-1)$ (angle). The normalized diameter of the UCA is equal to $\ell^{\text{UCA}} = 2\ell+2\Delta=2(Q+1)\Delta$.
According to Jake's model, the $({\texttt{i}_{n,q},\texttt{i}_{n',q'}})$-th entry of the normalized spatial correlation matrix is, for any user-to-AP link, given by
\begin{align}
    \!\![\mathbf{J}^{\text{UCA}}]_{\texttt{i}_{n,q} ,\texttt{i}_{n',q'} } \!=\!j_0\!\left(\!2\pi\sqrt{\!\left(\frac{\Delta_z \ell^{\text{UCA}}}{2(Q\!+\!1)}\right)^{\!\!2} \!\!+\! \left(\frac{\Delta_y \ell^{\text{UCA}}}{2(Q\!+\!1)}\right)^{\!\!2}}\right)\!,\!\!
\end{align}
where 
\begin{align*}
  \Delta_z \!&=\!\left|{\left(\!q \!+\! \frac{1}{2}\right)\!\cos\!\left(\frac{2\pi(n\!-\!1)}{N}\right) \!-\! \left(\!q' \!+\! \frac{1}{2}\right)\!\cos\!\left(\frac{2\pi(n'\!-\!1)}{N}\right)\!}\right|,\! \\
  \Delta_y \!&=\!\left|{\left(\!q \!+\! \frac{1} {2}\right)\!\sin\!\left(\frac{2\pi(n\!-\!1)}{N}\right) \!-\! \left(\!q' \!+\! \frac{1}{2}\right)\!\sin\!\left(\frac{2\pi(n'\!-\!1)}{N}\right)}\right|.\!
\end{align*}
These values are readily obtained by transforming the distances between the center positions of the two ports from the polar to the Cartesian $yz$-plane. 

The alternative representation for the normalized spatial correlation matrix based on the local scattering model can be obtained via~\eqref{eq:J:UPA:Jakes} upon converting the antenna port distances $\dH$ and $\dV$ from the polar to the Cartesian coordinates.     

\section{Proposed Uplink Training Scheme}
\label{estimation_sec}

Uplink training is particularly challenging in FA-based CF-mMIMO systems, primarily due to the dynamic and location-dependent nature of the FA ports. A conventional CF-mMIMO AP has a fixed number of antennas with fixed known spatial positions. The channel between each user and each antenna is fairly stable in structure, so the AP only needs to estimate a relatively well-defined set of channel coefficients. Once trained, the channel model remains valid unless significant changes occur in mobility or fading.

On the other hand, an FA can reconfigure its port location, meaning that each AP may have to estimate the channel for many possible antenna positions, specifically $NQ$ locations, in order to identify the optimal placement for the FA port that ensures the best performance gains. This increases pilot overhead, computational complexity, and channel variability, compared to conventional CF-mMIMO with fixed antenna elements. Additionally, uplink training becomes more challenging in the presence of NLoS channel components\footnote{In an LoS propagation environment, the channel can be easily estimated by using geometric principles~\cite{hu2024fluid}.}. However, the channel responses across nearby antenna ports are spatially correlated; thus, this correlation can be exploited to reduce the training overhead by training only a selected subset of antenna positions and inferring the channel at other positions, thereby balancing the number of probed port positions and the pilot symbols per position. 

\subsection{Pilot-based LMMSE Channel Estimation}
Let us assume that the uplink training employs $\tau_p$ channel uses, in which each user transmits a $\tau_p$-long pilot sequence. Let $\bm{\phi}_{t_k} \in \mathbb{C}^{\tau_p}$ be the pilot with index $t_k$, transmitted by user $k$. Every pilot is drawn from a set a $\tau_p$ mutually orthogonal sequences, such that    
\begin{align}
\bm{\phi}_{t_k}\trans\!\bm{\phi}_{t_j}^\ast =
\begin{cases} 
\tau_p, & \text{if } t_k = t_j, \\
0, & \text{otherwise}.
\end{cases}
\end{align}

In the proposed uplink training model characterization, we assume that any FA within an AP may potentially vary its port position over the training phase, namely, there may be either one or multiple port switching during the reception of the $\tau_p$ pilot-training symbols.
At the end of the training phase, the pilot signal matrix received at the $l$ AP is modeled as
\begin{align}
    \mathbf{Y}_{l} &= \sum_{j=1}^K \sqrt{ {\eta}_{\text{p},j}}[\bP_{\text{p},l}(1),\ldots, \bP_{\text{p},l}(\tau_p)] [\mathbf{I}_{\tau_p} \otimes \mathbf{h}_{j,l}]\diag (\bm{\phi}_{t_j}) \nonumber \\
    &\quad+ [\bP_{\text{p},l}(1)[\mathbf{Z}_l]_{:,1},\ldots, \bP_{\text{p},l}(\tau_p) [\mathbf{Z}_l]_{:,\tau_p}] \in \mathbb{C}^{N \times \tau_p},
    \label{Ylu}
\end{align}
where $\eta_{\text{p},j}$ is the pilot transmit power of user $j$, and $\mathbf{Z}_l \in \mathbb{C}^{NQ \times \tau_p}$ is the matrix of the additive noise at the receiver, with entries being independent and identically distributed, and each entry having mean of zero and variance of $\sigma^2$. The received pilot signal vector at the $u$-th channel use, $[{\mathbf{Y}}_{l}]_{:,u} \!\triangleq\! \mathbf{y}_{l,u} \in \mathbb{C}^{N}$, $u=1,\ldots,\tau_p$, is given by\footnote{In the notation $[{\mathbf{Y}}_{l}]_{:,u}$ and $[{\mathbf{Z}}_{l}]_{:,u}$, the index $u$ refers to the $u$-th column of the corresponding matrices, whereas in ${\mathbf{P}}_{\text{p},l}(u) \in \mathbb{C}^{N \times NQ}$, $u$ indicates the $u$-th instance of the port-location matrix when the $u$-th pilot symbol, $[\bm{\phi}_{t_k}\trans]_{u}$, is received.}
\begin{align}
\mathbf{y}_{l,u} = \sum_{j=1}^K \sqrt{\eta_{\text{p},j}} \mathbf{P}_{\text{p},l}(u) \mathbf{h}_{j,l} [\bm{\phi}_{t_j}\trans]_{u}+ \mathbf{P}_{\text{p},l}(u)[\mathbf{Z}_l]_{:,u}\,.
\label{ylu}
\end{align}
By stacking~\eqref{ylu} for all the values of $u$ into a vector, we obtain
\begin{align}
    \mathbf{\tilde y}_{l} = \left[\mathbf{y}\trans_{l,1}\, \mathbf{y}\trans_{l,2}\,,\cdots,\mathbf{y}\trans_{l,\tau_p}\right]\trans
=  \sum^K_{j=1} \sqrt{\eta_{\text{p},j}} {\bA}_{j,l}\bh_{j,l} + \tilde \bz_l\,,
\end{align}
where $\bA_{j,l} \in \mathbb{C}^{N\tau_p \times NQ}$ and $\tilde{\mathbf{z}}_l \in \mathbb{C}^{N\tau_p}$ are given by 
\begin{align}
    \bA_{j,l} = 
        \begin{bmatrix}
            [\bm{\phi}_{t_j}]_1{\bP}_{\text{p},l}(1)\\
            [\bm{\phi}_{t_j}]_2{\bP}_{\text{p},l}(2) \\[.25ex]
            \vdots \\[.25ex]
            [\bm{\phi}_{t_j}]_{\tau_p}{\bP}_{\text{p},l}(\tau_p)
        \end{bmatrix},\quad
    \tilde{\mathbf{z}}_l = 
        \begin{bmatrix}
            {\bP}_{\text{p},l}(1) [\mathbf{Z}_{l}]_1 \\
            {\bP}_{\text{p},l}(2) [\mathbf{Z}_{l}]_2 \\[.25ex] 
            \vdots \\[.25ex]
            {\bP}_{\text{p},l}(\tau_p) [\mathbf{Z}_{l}]_{\tau_p} 
        \end{bmatrix}.
        \label{Akl}
\end{align}
Assuming that the channel statistics are available at every AP, then the LMMSE estimation method can be used to estimate the channel of the $k$-th user as
\begin{align}
 &  {\mathbf{\hat h}}_{k,l}= \EX {  \mathbf{h}_{k,l}  |{ \mathbf{\tilde y}}_{l} } =  \sqrt{\eta_{\text{p},k}} \bR_{k,l} {\bA}\herm_{k,l}\mathbf{\Psi}_l^{-1} {\mathbf{\tilde y}}_{l}  \in \mathbb{C}^{NQ}\,,
\end{align}
where
\begin{align}
 \mathbf{\Psi}_l &= \sum\nolimits_{j=1}^{K} {\eta}_{\text{p},j} {\bA}_{j,l} \bR_{j,l} {\bA}_{j,l}\herm  + \sigma^2 \mathbf{I}_{N\tau_p}\,. \label{phi_l}
\end{align}
Hence, the collective vector of the estimates is distributed as
\begin{align}
{\mathbf{\hat h}_{k,l}} \sim \mathcal{CN}( 
\mathbf{0}_{NQ}, \eta_{\text{p},k} \mathbf{R}_{k,l} {\mathbf{A}_{k,l}\herm} \mathbf{\Psi}_{l}^{-1} \mathbf{A}_{k,l} \mathbf{R}_{k,l})\,,
\end{align}
and the corresponding uncorrelated estimation error, $\mathbf{\tilde{h}}_{k,l} = \mathbf{{h}}_{k,l} - \mathbf{\hat{h}}_{k,l}$, is distributed as $\mathbf{\tilde{h}}_{k,l} \sim \mathcal{CN}(\mathbf{0}_{NQ}, \mathbf{C}_{k,l})$, with 
\begin{align}
    \mathbf{C}_{k,l} = \mathbf{R}_{k,l} -  {\eta_{\text{p},k}} \mathbf{R}_{k,l} {\mathbf{A}_{k,l}\herm}  \mathbf{\Psi}_{l}^{-1} {\mathbf{A}_{k,l}}\mathbf{R}_{k,l}. 
    \label{Ckl}
\end{align}
\begin{remark}[Special case: Fixed port position]
If there is no port switching during the pilot transmission, only the channels towards $N$ ports are estimated. In this special case, we have
${\bP}_{\text{p},l}(u) = {\bP}_{\text{p},l}(\upsilon)$ for all $\upsilon \in \{1, \ldots, \tau_p\} \setminus \{u\}$, and thus the index $u$ can be omitted. Consequently, the LMMSE channel estimate is given by
\begin{align}
    {\mathbf{\hat h}}_{k,l} = \sqrt{\tau_p \eta_{\text{p},k}} \mathbf{R}_{k,l} {\bP}_{\text{p},l} \herm {{\mathbf{\Psi}}^{-1}_{l,t_k}} {\mathbf{\dot{y}}}_{l,t_k}\,,
    \label{hhat_short}
\end{align}
where $\mathbf{\dot{y}}_{l,t_k}$ is obtained from \eqref{ylu} by dropping $u$ and pre-multiplying by $\bm{\phi}_{t_k}^*/\sqrt{\tau_p}$, 
\begin{align}
    \mathbf{{\Psi}}_{l,t_k}  =  \sum\nolimits_{i \in \mathcal{P}_k} {\tau_p \eta_{\text{p},i}} {\bP}_{\text{p},l} \mathbf{R}_{i,l} {{\bP}\herm_{\text{p},l}} +\sigma^2 {\bP}_{\text{p},l} {\bP\herm_{\text{p},l}}, \label{psi2}
\end{align}
and the correlation matrix of the estimation error is given by
\begin{align}
\mathbf{C}_{k,l} = \mathbf{R}_{k,l} - \tau_p \eta_{\text{p},k} \, 
\mathbf{R}_{k,l} {\bP \herm_{\text{p},l} }  {\mathbf{\Psi}^{-1} _{l,t_k}}{\bP}_{\text{p},l} \mathbf{R}_{k,l}. \label{C_fix}
\end{align}
\end{remark}

\subsection{Optimal Selection of the Pilot-receiving Antenna Port}

The choice of the pilot-receiving port matrix $\{{\bP}_{\text{p},l}(u)\}$ plays a crucial role in the estimation process. If $\tau_p \geq Q$, potentially the channel towards every port can be estimated by letting each AP switch its antenna ports, in each pilot symbol transmission interval, in a round-robin manner. We refer to this strategy as the round-robin port switching (RR-PS).
Alternatively, a naive approach would involve letting the AP randomly switch the antenna ports during the pilot symbol transmission intervals. We refer to this baseline strategy as the random port switching (R-PS).

In this paper, we aim to optimize the pilot-receiving antenna port matrix for each AP $l$, $\{{\bP}_{\text{p},l}(u)\}$, to minimize the  per-AP NMSE of the channel estimates, which is defined as
\begin{align}
\text{NMSE}_{l} = \frac{1}{|\mathcal{K}_l|}\sum\limits_{k \in \mathcal{K}_l} \text{NMSE}_{k,l}\,.\label{nmsel}
\end{align}
where $\text{NMSE}_{k,l}$ is the \textit{statistical} NMSE of the channel estimate $\mathbf{\hat{h}}_{k,l}$ given by
\begin{align}
\text{NMSE}_{k,l} &=\frac{\E\{\mathbf{\tilde h}_{k,l}\mathbf{\tilde h}_{k,l}\herm\}}{\E\{\mathbf{h}_{k,l}\mathbf{h}_{k,l}\herm\}}
= \frac{\tr\{  \mathbf{C}_{k,l} \}}{\tr  \{ \mathbf{R}_{k,l}\}} \nonumber \\
&= 1-\frac{\eta_{\text{p},k} \tr\{\mathbf{R}_{k,l}\mathbf{A}_{k,l}\herm  \mathbf{\Psi}_{l}^{-1}\mathbf{A}_{k,l}\mathbf{R}_{k,l}\}}{\tr\{ \mathbf{R}_{k,l}\}} \label{nmsekl}\,.
\end{align}
This metric captures the accuracy of the channel estimation process performed at each AP. The corresponding optimization problem is formulated as
\begin{align}	
  \mathop {\text{minimize}}\limits_{\{{\bP}_{\text{p},l}(u)\}} & \quad \text{NMSE}_l\big(\{{\bP}_{\text{p},l}(u)\}_{u=1}^{\tau_p}\big)  
  \label{prob:NMSE} \\ 	
  \text{s.t.} &\quad {\bP}_{\text{p},l}(u)\in \mathcal{P}, \;\; \forall l, \forall u,  \label{prob:NMSE:c1} 
\end{align}
where $\mathcal{P}$ is the feasible set of the pilot-receiving port matrix.
The optimization problem in~\eqref{prob:NMSE} is, in general, \emph{non-convex}. First, each pilot-receiving port matrix $\bP_{\text{p},l}(u)$ is a binary selection matrix, and it belongs to a discrete feasible set $\mathcal{P}$. Hence, the feasible region of~\eqref{prob:NMSE} is non-convex by construction. Furthermore, the NMSE expression in~\eqref{nmsekl} depends on the matrices $\bP_{\text{p},l}(u)$ through both $\bA_{k,l}$ and the covariance matrix $\mathbf{\Psi}_l$, where $\mathbf{\Psi}_l^{-1}$ appears inside a matrix trace. The mapping $\bP_{\text{p},l}(u) \mapsto \bA_{k,l}\herm \mathbf{\Psi}_l^{-1} \bA_{k,l}$ is highly nonlinear, and the matrix inversion operation prevents the objective from being convex even under a continuous relaxation of $\bP_{\text{p},l}(u)$. Therefore,~\eqref{prob:NMSE} constitutes a combinatorial, non-convex optimization problem, and must be addressed using either discrete search strategies, heuristic algorithms, or suitable relaxation techniques. In this regard, we propose a heuristic algorithm, referred to as \textit{local NMSE-descent port selection} (LND-PS) described by the pseudo-code of Algorithm~\ref{Alg:optP}, that finds a local optimal set $\{{\bP}_{\text{p},l}(u)\}$ via an iterative local search strategy. Starting from an initial feasible configuration $\{\bP_{\text{p},l}^{(0)}(u)\}$, the algorithm sequentially explores neighboring configurations by modifying one pilot-receiving port matrix at a time while keeping the others fixed. A candidate update is accepted only if it results in a strictly lower NMSE value. This ensures that the NMSE decreases monotonically across iterations and, since the NMSE is bounded below, the algorithm is guaranteed to converge to a fixed point. However, due to the non-convexity of the objective, the obtained solution corresponds to a local rather than a global minimum, and the final result may depend on the initialization and the update sequence. 

\begin{remark}[Initialization strategy] The convergence point of the LND-PS algorithm depends on the initialization of the set $\{\mathbf{P}_{\text{p},l}^{(0)}(u)\}$. To ensure favorable performance and fast convergence, we recommend a statistically informed initialization. When $\tau_p \geq Q$, a RR-PS initialization provides uniform pilot exposure across all FA ports. When $\tau_p < Q$, although rare in practice, a covariance-aware initialization is preferred. Since the per-port average power is identical across ports (as dictated by the model in~\eqref{RJ}), port selection must be driven by the off-diagonal correlation entries. In this regard, a desirable solution is to maximize diversity among the selected ports so that the sampled channel directions are as ``statistically uncorrelated'' (informative) as possible; and to avoid redundancy caused by selecting strongly correlated ports that provide nearly the same observation. These goals directly translate into a standard information-theoretic criterion: \emph{maximize the log-determinant} (log-det) of the aggregated covariance of the chosen ports. Specifically, for each AP $l$, we determine the set of the initial active ports as
\begin{align}
\mathcal{S}^{\star} = \argmax\limits_{|\mathcal{S}| = N} \log \det\, \bR_l^{\rm {agg}}(\mathcal{S},\mathcal{S})\,,
\label{eq:init-rule}
\end{align}
where $\mathcal{S} \subset \{1,\ldots,NQ\}$, with $|\mathcal{S}|=N$, and $\bR_l^{\rm {agg}}(\mathcal{S},\mathcal{S})$ denotes the principal submatrix of the aggregated covariance matrix, $\bR_l^{\rm {agg}} = \sum\nolimits_{k \in \mathcal{K}_l} \bR_{k,l}$, indexed by $\mathcal{S}$.
Hence, the pilot-receiving port matrix is initialized, for all $u=1,\cdots,\tau_p$, as
\begin{align}
    [\mathbf{P}_{\text{p},l}^{(0)}(u)]_{n,\texttt{i}_{n,q}}=
    \begin{cases}
        1,&\forall n,\,\forall q: \texttt{i}_{n,q} \in \mathcal{S}\,,  \\[.5ex]
        0,&\text{otherwise}\,.
    \end{cases}
\end{align}
Note that the port selected according to the rule in~\eqref{eq:init-rule} holds for all the pilot symbols $u=1,\ldots,\tau_p$, as the spatial correlation matrices $\{\bR_{k,l}\}$ do not vary within a channel coherence block.
This log-det initialization yields a set of ports that are collectively informative and typically accelerates the subsequent LND-PS iterations compared to purely random or RR-PS initializations.
\end{remark}
Despite this limitation, the proposed approach is computationally efficient, easy to implement at each AP, and provides a practical trade-off between performance and complexity, making it suitable for real-time port-selection control.
\begin{remark}[Computational complexity]
For each AP $l$, one iteration of the LND-PS algorithm evaluates candidate updates for $\tau_p$ pilot symbols, and for each symbol, there are $Q$ possible FA-port choices for each antenna. Each NMSE evaluation requires constructing the matrices $\mathbf{A}_{k,l}$ and $\mathbf{\Psi}_l$, and computing a matrix inverse of size $N\tau_p \times N\tau_p$. Therefore, the per-iteration computational complexity for each AP is
\begin{align}
\mathcal{O}\left(\tau_p NQ\Big((N\tau_p)^3 + K (N\tau_p)^2\Big)\right).
\end{align}
Since the maximum number of iterations is $I_{\max}$, the overall complexity per AP is
\begin{align}
\mathcal{O}\left(I_{\max} \tau_p NQ(N\tau_p)^3\right).
\end{align}
This is significantly more efficient than exhaustive search, which scales as $\lvert \mathcal{P} \rvert^{\tau_p}$ per AP and is therefore computationally infeasible for moderate values of $\tau_p$ or $Q$. Moreover, if $\tau_p$ is an integer multiple of $Q$, then the port-switching rate can be reduced by a factor $\tau_p/Q$, i.e., from switching one port per channel use to switching one port every $\tau_p/Q$ channel uses. This, in turn, results in a reduced computational complexity:
\begin{align}
\mathcal{O}\!\left(\!I_{\max} \frac{\tau_p}{\tau_p/Q} NQ(N\tau_p)^3\!\right)\! \rightarrow \!\mathcal{O}(I_{\max} NQ^2(N\tau_p)^3).
\end{align}
\end{remark}
\begin{algorithm}[t]
\caption{LND-PS: Local NMSE-Descent Port Selection}
\label{Alg:optP}
\begin{algorithmic}[1] \small
\State \textbf{Input:} Initial $\{{\bP}^{(0)}_{\text{p},l}(u)\}$, $I_{\max}$, $\epsilon$
\For{$l = 1,\ldots,L$}
    \State $i \gets 0$; $J_l^{(0)} \gets \text{NMSE}_l(\{{\bP}^{(0)}_{\text{p},l}(u)\})$
    \While{$i < I_{\max}$}
        \For{$u = 1,\ldots,\tau_p$}
            \For{$n = 1,\ldots,N$}
                \For{$q = 1,\ldots,Q$}
                \State Set ${\bP}^{\star}_{\text{p},l}(u)$ by changing one port position
                \State Form $\{{\bP}_{\text{p},l}(u)\}$ by substituting ${\bP}^{\star}_{\text{p},l}(u)$
                \State Compute $J_l^{\star} \gets \text{NMSE}_l(\{{\bP}_{\text{p},l}(u)\})$
                \If{$J_l^{\star} < J_l^{(i)}$}
                    \State ${\bP}^{(i+1)}_{\text{p},l}(u) \gets {\bP}^{\star}_{\text{p},l}(u)$; $J_l^{(i+1)} \gets J_l^*$
                \EndIf
                \EndFor
            \EndFor
        \EndFor
        \If{$|J_l^{(i+1)} - J_l^{(i)}| < \epsilon$} \textbf{break} \EndIf
        \State $i \gets i + 1$
    \EndWhile
\EndFor
\State \textbf{Output:} Optimized $\{{\bP}_{\text{p},l}(u)\}$ for all APs
\end{algorithmic}
\end{algorithm}

\section{Uplink Performance Analysis}\label{metric}
In this section, we derive achievable SE expressions for the FA-empowered CF-mMIMO uplink. Both centralized and distributed signal processing are considered.

Let us denote the data-receiving antenna port matrix at AP $l$ as $\bP_{\text{d},l}$. Moreover, let $s_k$, $\E\{|s_k|^2\} = 1,\forall k$, $\E\{s_ks_j^{\ast}\}=0, k\neq j$, be the data symbol of the $k$-th user, and $\eta_{\text{d},k}$ be the transmit power of user $k$. 
The $N$-dimensional uplink data signal received at the $l$-th AP is given by:
\begin{align}
    \mathbf{r}_l = \sum\nolimits_{k=1}^K \sqrt{\eta_{\text{d},k}} \,\bP_{\text{d},l}\mathbf{h}_{k,l} s_k + \tilde{\mathbf{n}}_l\,,
    \label{eq:y_m}
\end{align}
with $\tilde{\mathbf{n}}_l \sim \mathcal{CN}(0,\sigma^2 \mathbf{I}_N)$ being the additive thermal noise. 
To compute a soft estimate of the data transmitted by the $k$-th user, the uplink data signal in~\eqref{eq:y_m} is linearly combined with a proper combining vector, $\mathbf{v}_{k,l} \in \mathbb{C}^N$, to give $r_{k,l} = \mathbf{v}_{k,l}\herm\mathbf{r}_{l}$. By letting $\varrho_{k,j,l} = \mathbf{v}_{k,l}\herm\bP_{\text{d},l}\mathbf{h}_{j,l}$, we thus have, 
\begin{align}
	r_{k,l} &=\sqrt{\eta_{\text{d},k}}\varrho_{k,k,l} s_k + \!\!\sum^K_{j=1,\,j\neq k}\!\!\sqrt{\eta_{\text{d},j}}\varrho_{k,j,l} s_j + \bv_{k,l}\herm \tilde{\mathbf{n}}_l\,.
    \label{eq:y_km}
\end{align}

\subsection{Centralized Processing}
In the uplink centralized operation, the CPU, based on the channel estimates. designs the combining vectors accordingly. The uplink data signals received at the APs are then sent to the CPU which computes $r_{k,l}$ as in~\eqref{eq:y_km}, for all the APs serving user $k$, namely $\forall l \in \mathcal{K}_l$, and jointly combines them to obtain the estimate of $s_k$, that is $\widehat{s}_k = \sum_{l\in \mathcal{L}_k} r_{k,l}$.
By applying the same information-theoretic tools of \cite[Theorem 5.1]{demir2021foundations} on the signal model of $\widehat{s}_k$, an achievable SE for the $k$-th user, expressed in [bit/s/Hz], is given by 
\begin{equation}
    \mathsf{SE}_k^{\rm c} = \left(1-\tau_p/\tau_c\right) \EX{\log_2\left(1+\mathsf{SINR}_k^{\rm c}\right)}\,,
    \label{eq:SE:centralized}
\end{equation}
where $\tau_c$ is the length of the coherence block, and the effective SINR, upon letting $\hat{\varrho}_{k,j,l} = \mathbf{v}_{k,l}\herm\bP_{\text{d},l}\hat{\mathbf{h}}_{j,l}$, is given by
\begin{align}
    \mathsf{SINR}_k^{\rm c} \!=\! \frac{\eta_{\text{d},k}\left|\sum\nolimits_{l \in \mathcal{L}_k} \hat{\varrho}_{k,k,l}\right|^2}{\!\sum\limits^K_{{j=1,\,j\neq k}}\!\eta_{\text{d},j}\left|\sum\nolimits_{l \in \mathcal{L}_k}  \hat{\varrho}_{k,j,l}\right|^2+\varsigma_k}\,,
    \label{sinr_up} 
\end{align}
and with
\begin{align}
 \!\varsigma_k \!=\!  \sum\limits_{l \in \mathcal{L}_k} \mathbf{v}_{k,l}\herm\!\left( 
 \sum\nolimits_{j=1}^K {\eta_{\text{d},j}}\bP_{\text{d},l}\bC_{j,l} \bP\trans_{\text{d},l} \!+\! \sigma^2 \bI_{N}
 \right)\!\mathbf{v}_{k,l}.
\end{align}
Two centralized combining schemes are considered herein: $(i)$ MRC and $(ii)$ \textit{partial MMSE} (P-MMSE). 
MRC constitutes the combining scheme with the lowest complexity, and consists of setting the combining vector, intended for the link between AP $l$ and user $k$, as
\begin{align} 
    \mathbf{v}_{k,l}^{\text{MRC}} = \bP_{\text{d},l}\hat{\mathbf{h}}_{k,l}. \label{MR}
\end{align}
P-MMSE combining serves as a scalable and computationally efficient 
approximation of the optimal MMSE combining scheme~\cite[Sec.~5.1.4]{demir2021foundations} and consists of setting the collective combining vector intended for user $k$ as
\begin{align}
    \!\bv_{k}^{\text{P-MMSE}} \!=\! {\eta}_{\text{d},k}\! \left(\sum_{i\in\mathcal{S}_k}{\eta}_{\text{d},i}\bD_k\bB_{i}\bD_k\herm \!+\! \sigma^2\bI_{LN}\! \right)^{\!\!-1}\!\!\! \bD_k {\bP}_{\text{d}}\hat{\bh}_k,\! 
    \label{PMMSE}
\end{align}
with $\bB_{i} = {\bP}_{\text{d}} \hat\bh_{i} \hat\bh_{i}\herm \bP_{\text{d}}\herm + \bP_{\text{d}} \bC_{i} \bP_{\text{d}}\herm$, and 
\begin{align}
    \hat\bh_k &\triangleq [\hat\bh_{k,1}\trans, \ldots, \hat\bh\trans_{k,L}] \in \mathbb{C}^{LNQ}\,, \\
    \mathbf{P}_{\text{d}} &\triangleq \diag(\mathbf{P}_{\text{d},1}, \dots, \mathbf{P}_{\text{d},L}) \in \mathbb{C}^{LN \times LNQ}\,,\\
    \mathbf{C}_i &\triangleq \diag(\mathbf{C}_{i,1}, \dots, \mathbf{C}_{i,L}) \in \mathbb{C}^{LNQ \times LNQ}\,,\\ 
    \bD_k &\triangleq \diag\big(\vartheta_1 \mathbf I_N, \dots, \vartheta_L \mathbf I_N \big) \in \mathbb{C}^{LN \times LN}
\end{align}
and $\vartheta_\ell = 1$ if $\ell \in \mathcal{L}_k$, or 0 otherwise, while $\mathcal{S}_k = \{j\,:\, \mathcal{L}_k \cap \mathcal{L}_j \neq \emptyset\}$ denotes the set of users sharing at least one serving AP with user $k$.
The combining vector intended for the link between AP $l$ and user $k$ is obtained from the collective vector as $\bv_{k,l}^{\text{P-MMSE}} = [\bv_k^{\text{P-MMSE}}]_{(l-1)N + 1 : l N}$.

An alternative expression for an achievable SE under MRC operation can be derived by leveraging the \textit{use-and-then-forget} (UatF) capacity-bounding technique~\cite{redbook}, which assumes that the CPU uses its knowledge of the estimates only for designing the combining vectors as in~\eqref{MR}, while forgetting that in the data decoding by treating the channel as deterministic. It provides a pessimistic evaluation of the achievable SE when the channel undergoes significant fluctuations.
By using the same methodology as in~~\cite[Sec.~5.1.2]{demir2021foundations}, the resulting effective SINR, denoted as \(\text{SINR}^{\text{UatF}}_k\), the UatF lower bound on the capacity is given by
\begin{equation}
    \mathsf{SE}_k^{\text{UatF}} = \left(1-\tau_p/\tau_c\right) \log_2\left(1+\mathsf{SINR}^{\text{UatF}}_k\right)\,,
    \label{eq:SE:centralized:UatF}
\end{equation}
where $\mathsf{SINR}^{\text{UatF}}_k$ is given in~\eqref{UatF} at the top of the next page.
\begin{figure*}
\begin{align}
    \text{SINR}^{\text{UatF}}_k \!=\! \frac{\eta_{\text{d},k}\left|\sum\nolimits_{l \in \mathcal{L}_k} \E\{\bv_{k,l}\herm{\bP_{\text{d},l}}\bh_{k,l}\}\right|^2}
    {\sum\nolimits_{j=1}^K \eta_{\text{d},j} \EX{\big|\sum\nolimits_{l \in \mathcal{L}_k} \bv_{k,l}\herm \bP_{\text{d},l} \bh_{j,l}\big|^2} \!-\! \eta_{\text{d},k}\left|\sum\nolimits_{l \in \mathcal{L}_k} \E\{\bv_{k,l}\herm \bP_{\text{d},l}\bh_{k,l}\}\right|^2 \!+\! \sigma^2 \sum\nolimits_{l \in \mathcal{L}_k} \EX{\norm{\bv_{k,l}}^2}}
    \label{UatF} 
\end{align}
\hrulefill
\begin{align}
    \text{SINR}^{\text{UatF}}_k \!=\! \frac{\eta_{\text{d},k}\left|\sum\nolimits_{l\in\mathcal{L}_k}\tr\{\bP_{\text{d},l}\boldsymbol{\Gamma}_{k,l} \bP\herm_{\text{d},l}\}\right|^2}{\sum\limits_{j=1}^K \sum\limits_{l\in\mathcal{L}_k} \eta_{\text{d},j}\tr\{\bP_{\text{d},l}\bR_{j,l} \bP\herm_{\text{d},l} \bP_{\text{d},l}\boldsymbol{\Gamma}_{k,l} \bP\herm_{\text{d},l} \} \!-\! \eta_{\text{d},k}\left|\sum\nolimits_{l\in\mathcal{L}_k}\tr\bP_{\text{d},l}\boldsymbol{\Gamma}_{k,l} \bP\herm_{\text{d},l}\}\right|^2 \!+\! \sigma^2\sum\limits_{l\in\mathcal{L}_k}\tr\{\bP_{\text{d},l}\boldsymbol{\Gamma}_{k,l}\bP\herm_{\text{d},l}\}}
    \label{eq:ClosedFormUatF}    
\end{align}
\hrulefill
\end{figure*}
\begin{proposition}
A closed-form expression for $\text{SINR}_k^{\text{UatF}}$ in~\eqref{UatF}, under the assumption of spatially correlated Rayleigh fading, LMMSE channel estimation, no pilot reuse, i.e., $\tau_p \geq K$, and MRC scheme, is given in~\eqref{eq:ClosedFormUatF} at the top of the next page, where $\mathbf{\Gamma}_{k,l}$ is defined as
\begin{equation}
    \boldsymbol{\Gamma}_{k,l} =  \eta_{\text{p},k} \mathbf{R}_{k,l} {\mathbf{A}_{k,l}\herm}  \mathbf{\Psi}_{l}^{-1} {\mathbf{A}_{k,l}}\mathbf{R}_{k,l},
    \label{Gammakl2}
\end{equation}
and ${\mathbf{A}_{k,l}\herm} $ and $\mathbf{\Psi}_{l}$ are defined in \eqref{Akl} and \eqref{phi_l}, respectively.
\end{proposition}
\begin{IEEEproof}
The proof is provided in Appendix A.
\end{IEEEproof}

\subsection{Distributed Processing}

In the uplink distributed operation, channel estimation and receive combining are performed locally at each AP. The local estimates resulting from this combining operation are then sent to the CPU, which further linearly combines them as
\begin{equation}
    \widehat{s}_k= \sum\nolimits_{l\in \mathcal{L}_k} \alpha^{*}_{k,l} r_{k,l}\,,
    \label{eq:softestimate_dist}
\end{equation}
where $\alpha_{k,l}$ is the deterministic weight assigned by the CPU to the local estimate of the signal from user $k$ and processed by AP $l$. Accordingly, an achievable SE of user $k$ in the distributed processing is given by~\cite[Sec. 5.2.1]{demir2021foundations}
\begin{equation}
    \mathsf{SE}_k^{\rm d} = \left(1-\tau_p/\tau_c\right) \EX{\log_2\left(1+\mathsf{SINR}_k^{\rm d}\right)}\,,
    \label{eq:SE:distributed}
\end{equation}
where the effective SINR is
\begin{align}
    \text{SINR}_k^{\rm d} = \frac{\eta_{\text{d},k}|\boldsymbol{\alpha}_k\herm \EX{\bg_{kk}}|^2}{\boldsymbol{\alpha}_k\herm \left(\bG_k +\bF_k \right)\boldsymbol{\alpha}_k}\,,
    \label{SINRd}
\end{align}
with $\bG_k \!=\! \sum_{j=1}^K {\eta_{\text{d},j}}\E\{\bg_{kj}\bg_{kj}\herm\} \!-\!\eta_{\text{d},k}\E\{\bg_{kk}\}\!\E\{\bg_{kk}\herm\}$, and $\bF_k \!=\! \sigma^2 \diag( \E\{\norm{\bD_{k,1}\bv_{k,1}}^2\},\ldots,\E\{\norm{\bD_{k,L}\bv_{k,L}}^2\})$. Also, $\bg_{kj}=\left[g_{kj,1},\ldots,g_{kj,L}\right]\trans$, where $g_{kj,l}\triangleq \bv_{k,l}\herm \bD_{k,l}\bP_{\text{d},l} \hat{\bh}_{j,l}$, and $\bD_{k,l}=\vartheta_l\bI_N$ with $\vartheta_l = 1$ if $l \in \mathcal{L}_k$, or 0 otherwise.
The SINR expression in~\eqref{SINRd} is ``scalably'' optimized by the so-called
\textit{nearly-optimal large-scale fading decoding (n-opt LSFD)} vector defined in~\cite[eq.~(5.41)]{demir2021foundations} as
\begin{align}
    \boldsymbol{\alpha}_{k} \!=\! {\eta}_{\text{d},k} \Bigg( \sum_{j\in \mathcal{S}_k} \eta_{\text{d},j} \E\{\bg_{kj}\bg_{kj}\herm\} \!+\! \bF_k \!+\! \tilde{\bD}_k \Bigg)^{\!\!-1}\!\! \E\{\bg_{kk}\}
\end{align}
where \({\tilde{\bD}_k} \in \mathbb{R}^{L\times L}\) is a regularization term, that is a diagonal matrix whose \((l,l)\)-th element is equal to one whenever \( l\notin \mathcal{L}_k \), and zero otherwise.
As for the local combining at the APs, we consider MRC and local partial MMSE (LP-MMSE). The combining vector designed by AP $l$ for the $k$-th user, is, for the MRC case, given by~\eqref{MR}. Whereas in the case of the LP-MMSE combining scheme, it is defined as~\cite[Section 5.2.2]{demir2021foundations},
\begin{align}   
    \bv_{k,l}^{\text{LP-MMSE}} \!=\! \eta_{\text{d},k}\! \left( \sum_{i\in\mathcal{K}_l} \eta_{\text{d},i} \bB_{i,l} \!+\! \sigma^2 \bI_N \right)^{\!\!-1}\!\!\bD_{k,l}{\bP}_{\text{d},l} \hat\bh_{k,l}
    \label{LPMMSE}
\end{align}
where $\bB_{i,l} = \bP_{\text{d},l}(\hat\bh_{i,l} \hat\bh_{i,l}\herm + \bC_{i,l})\bP_{\text{d},l}\trans$.

\subsection{Antenna Port Selection for Sum-SE Maximization}
We next present an iterative algorithm that sub-optimally selects the antenna ports maximizing the sum-SE. The optimization problem for the sum-SE maximization with respect to the data-receiving antenna port matrix is formulated as
\begin{subequations}
    \begin{align}
    \mathop {\text{max}} \limits_{\{{\bP}_{\text{d},l}\}} & \;
    \boldsymbol{\Sigma}_{\mathsf{SE}} = \sum\nolimits_{j=1}^K \text{SE}_j^{\text{lb}} (\{{\bP}_{\text{d},l}\})
    \label{eq:Q_subX} \\
	\text{s.t.} &\; [\bP_{\text{d},l}]_{n,m} \!\in\! \{ 0,1\},
            \; l=1,\ldots,L,~n=1,\ldots,N, \nonumber \\
            &\;\qquad m = (n-1)Q + 1, \ldots, nQ,\\
            &\; \norm{[\bP_{\text{d},l}]_{n,:}}_0 \!=\! 1,  \quad l=1,\ldots,L,~n=1,\ldots,N, \label{eq:constraints_positions}
    \end{align}%
\label{bb}%
\end{subequations}%
where the superscript $\text{lb} \in \{\text{c}, \text{d}, \text{UatF}\}$ refers to the capacity lower-bound expressions in~\eqref{eq:SE:centralized},~\eqref{eq:SE:centralized:UatF} and~\eqref{eq:SE:distributed}, respectively.
The optimization problem in~\eqref{bb} is a combinatorial assignment problem over a finite feasible set of data-receiving antenna port matrices, and can be tackled via iterative \textit{greedy} search. Specifically, we employ an alternating optimization (AO) scheme in which the port configuration of one AP is optimized at a time, with the others fixed, and iterate this procedure until convergence of the objective.
Algorithm~\ref{alg:antenna_position} details the iterative optimal selection of the antenna port at each AP to maximize the sum-SE.
The matrices $\{\bP_{\text{d},l}\}$, $l = 1,\ldots,L$, are arbitrarily initialized. The sum-SE attained by this initial configuration is denoted by $\boldsymbol{\Sigma}^{(0)}_{\mathsf{SE}}$. At each iteration, for an arbitrary AP $l$ and antenna $n$, the $m^{\star}$-th port that maximizes the sum-SE is selected, hence the corresponding entry of the data-receiving antenna port matrix, $[{\bP}_{\text{d},l}]_{n,m^{\star}}$, is set to $1$, whereas the other entries, $[{\bP}_{\text{d},l}]_{n,m}$, with $m \neq m^{\star}$, are set to $0$. This iterative procedure repeats until the relative change in the value of the objective~\eqref{eq:Q_subX} falls below the convergence threshold $\epsilon$ or the maximum number of iterations $i_{\max}$ is reached. The overall computational complexity of Algorithm~\ref{alg:antenna_position} is at most $\mathcal{O}(i_{\max}LQN)$, which is considerably smaller than the computational complexity of the optimal joint port selection across the $N$ antennas, that is $\mathcal{O}(LQ^N)$, and affordable for realistic values of $Q$ and $N$. 
{\begin{algorithm}[!t]
\caption{sum-SE maximization via port AO}\label{alg:antenna_position}
\begin{algorithmic}[1]\small
\State Initialize $\epsilon$, $i_{\max}$, $i=0$, STOP $=$ false, and $\{\bP_{\text{d},l}\}^L_{l=1}$ 
\State $\boldsymbol{\Sigma}^{(0)}_{\mathsf{SE}} \gets \sum\nolimits_{j=1}^K \text{SE}_j^{\text{lb}} (\{{\bP}_{\text{d},l}\})$
\While{STOP $=$ false}
	\State $i \gets i+1$; 
    \For{$l=1,\ldots,L$}
       \For{$n=1,\ldots,N$}
          \State $m^{\star} \gets \arg\max\limits_{m \in \{(n-1)Q+1, \ldots, nQ\}} \sum\nolimits_{j=1}^K \text{SE}^{\text{lb}}_j (\{{\bP}_{\text{d},l}\})$
        \State Set $[\bP_{\text{d},l}]_{n,m^{\star}} \gets 1$;
        \State Set $[\bP_{\text{d},l}]_{n,m} \gets 0$, $\forall m \neq m^{\star}$;
	\EndFor
 \EndFor
 \State $\boldsymbol{\Sigma}^{(i)}_{\mathsf{SE}} \gets \sum\nolimits_{j=1}^K \text{SE}_j^{\text{lb}} (\{{\bP}_{\text{d},l}\})$
\If{ $\big(\big|\boldsymbol{\Sigma}^{(i)}_{\mathsf{SE}}-\boldsymbol{\Sigma}^{(i-1)}_{\mathsf{SE}}\big|\big/\boldsymbol{\Sigma}^{(i-1)}_{\mathsf{SE}}\leq \epsilon\big)$ OR $\; (i\geq i_{\max}$)}
\State STOP $=$ true;
\EndIf
\EndWhile
\end{algorithmic}
\end{algorithm}
}
    
\section{Numerical Results}\label{numerical}
In this section, we present the simulation results to validate the effectiveness of the proposed port-selection algorithms. 
In our simulations, we adopt the settings reported in Table~\ref{table:param}, unless otherwise stated. Moreover, we set the convergence threshold for both the proposed algorithms as $\epsilon = 10^{-8}$. The coefficient $\beta_{k,l}$ captures the effects of both path loss and shadowing and herein is assumed to follow the 3GPP Urban Micro pathloss model as defined in~\cite[Table B.1.2.1-1]{3GPP_36814_model}:
\begin{align*}
[\beta_{k,l}]_{\text{dB}} = 22.7 + 36.7\log_{10}(d_{k,l}) + 26\log_{10}(f_{\text{c}}) + \tilde{F}_{k,l}, 
\end{align*}
where $d_{k,l}$ is the 3D distance in meters between AP $l$ and user $k$, and $f_{\text{c}}$ is the carrier frequency in GHz. 
Importantly, we reasonably assume that $\beta_{k,l}$ remains constant over the ports and the antenna elements within the $l$-th AP, hence the lack of dependency on the port and FA indices. Moreover, $\tilde{F}_{k,l}\sim \mathcal{N}(0,4^2)$ is the log-normal correlated shadowing whose correlation, across different APs and users, is characterized according to the model~\cite[Table B.1.2.2.1-4]{3GPP_36814_model}
\begin{align}
    \E\{\tilde{F}_{k,l}\tilde{F}_{i,j}\} = 
        \begin{cases}
            4^2 2^{-\delta_{ki}/9 \;\text{m}}  & l=j \\
            0 & l \ne j
        \end{cases}
    \label{F}
\end{align}
where $\delta_{ki}$ is the distance between users $k$ and $i$.
\begin{table}[!t]
\renewcommand{\arraystretch}{1.2}
\centering
\caption{Simulation Parameters}
\begin{tabular}{|c|c|c|c|c|}
\cline{1-2}\cline{4-5}
\textbf{Parameter} & \textbf{Value} & \hspace{-3.5mm} & \textbf{Parameter} & \textbf{Value} \\
\cline{1-2}\cline{4-5}
Carrier frequency, $f_c$ & 3.5 GHz & \hspace{-3.5mm} & $L$ & 64, 100\\
Bandwidth & 20 MHz & \hspace{-3.5mm} & $N$ & 4\\
Network area & 400m$\times$400m & \hspace{-3.5mm} & $Q$ & 5\\
$\tau_c$, $\tau_p$ & 200, $K$ samples & \hspace{-3.5mm} & $K$ & 10, 20\\
AP, UE height & 10 m, 1.65 m & \hspace{-3.5mm} & $|\mathcal{L}_k|$ & 5 $\forall k$\\
PSD noise & -174 dBm/Hz & \hspace{-3.5mm} &  $\eta_{\text{max}}$ & 100 mW \\ 
Noise figure & 7 dB & \hspace{-3.5mm} & $\Delta$ & 0.5\\
\cline{1-2}\cline{4-5}
\end{tabular}\label{table:param}
\vspace{-4mm}
\end{table}
As for the uplink power allocation strategy, we consider: $(i)$ \textit{uniform full power transmission} (UFP), in which all the users transmit their data with the available power budget, $\eta_{\text{max}}$; $(ii)$ \textit{max-min fairness} (MMF) power allocation, in which the power control coefficients are optimized to maximize the minimum SE according to~\cite[Algorithm 7.1]{demir2021foundations}; and $(iii)$ \textit{fractional power allocation} (FPA)~\cite{demir2021foundations,ScalabilityAspects}, in which the power control coefficient intended for user $k$ is set as
\begin{equation}
\eta_{\mathrm{d},k} = \eta_{\max} \frac{\left( \sum_{l \in \mathcal{L}_k} \beta_{k,l} \right)^{\varpi}}{\max\limits_{i \in \mathcal{S}_k} \left( \sum\nolimits_{l \in \mathcal{L}_i} \beta_{i,l} \right)^{\varpi}},
\end{equation}
where the exponent $\varpi$ defines the power allocation policy.

With regard to uplink channel estimation, beyond the R-PS, RR-PS, and the proposed LND-PS scheme, we consider two additional benchmarks: the \textit{fixed} port-switching (F-PS) scheme, in which the FA simply behaves as a ``conventional'' antenna, as the fluid does not move across the antenna ports during the uplink training; and the $\nu$\textit{-skip} scheme, proposed in~\cite{fe6}, in which the parameter $\nu$ specifies how many subsequent ports are bypassed in the uplink training so that the AP estimates the channel only at periodically spaced ports of each of its antennas rather than at every port, extending the standard RR-PS approach. As a result, if $\nu=0$, then the $\nu$\textit{-skip} and the RR-PS schemes coincide, whereas if $\nu=Q-1$, then the $\nu$\textit{-skip} scheme reduces to the F-PS approach.
\begin{figure}[t]
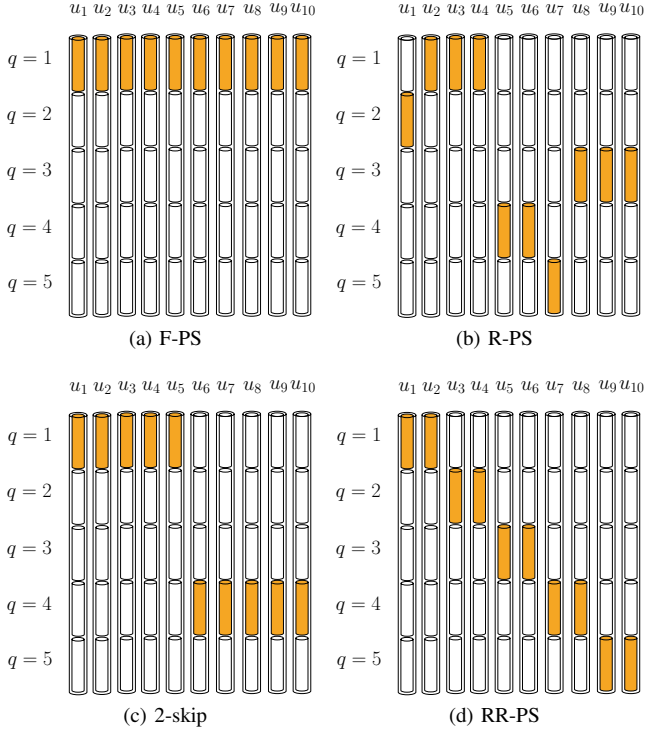

    \centering
    \subfloat[F-PS]{\resizebox{.48\columnwidth}{!}{\input{Fixed_PS}} \label{fig_config_sub1}}%
    \subfloat[R-PS]{\resizebox{.48\columnwidth}{!}{\input{R_PS}} \label{fig_config_sub2}}%
    \\
    \subfloat[2-skip]{\resizebox{.48\columnwidth}{!}{\input{2skip_PS}} \label{fig_config_sub3}}%
    \subfloat[RR-PS]{\resizebox{.48\columnwidth}{!}{\input{RR_PS}} \label{fig_config_sub4}}%
    \caption{Illustrative example of the operation of the pilot-switching schemes serving as a benchmark. Each colored block represents the active port selected for channel estimation over the corresponding pilot symbol. Settings: $Q=5$, $\nu=2$, and $\tau_p = 10$.}
    \label{fig_config}
\end{figure}
Fig.~\ref{fig_config} provides an illustrative example of the operation of the pilot-switching schemes serving as a benchmark for a single-antenna AP with $Q=5$, $\nu=2$, and $\tau_p = 10$.\footnote{If $\nu=2$ and $Q=5$, then each AP estimates, for all its antennas, the channels towards the ports with local indices in the set $\mathcal{S}_q=\{1,4\}$ over $\lfloor \tau_p/|\mathcal{S}_q|\rfloor$ channel uses each.}

\begin{figure}[t]	
    \centering
    \includegraphics[width=.95\columnwidth]{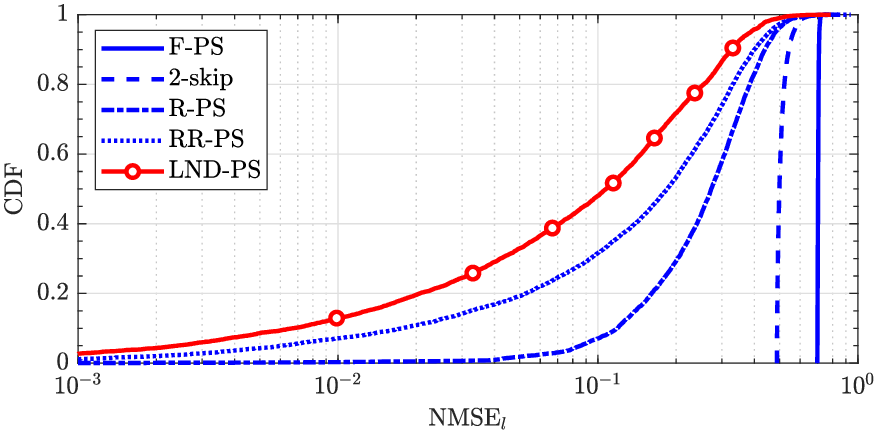}
    \caption{CDF of the per-AP NMSE for different port-switching schemes. Settings: $K=10$, $L=64$, and ULA-based APs.}
    \label{fig0101}
\end{figure}

Fig.~\ref{fig0101} illustrates the cumulative distribution function (CDF) of the per-AP NMSE in~\eqref{nmsel}, over several random realizations of network deployments, for different port-switching schemes. 
Fig.~\ref{fig0101} primarily highlights the clear advantage of our proposed LND-PS algorithm in significantly reducing the NMSE compared to the considered prior strategies. The RR-PS scheme outperforms the R-PS, $\nu$\textit{-skip}, with $\nu=2$, and F-PS schemes as it maximizes diversity among the selected active ports.

\begin{figure}[t]	
    \centering
    \includegraphics[width=.95\columnwidth]{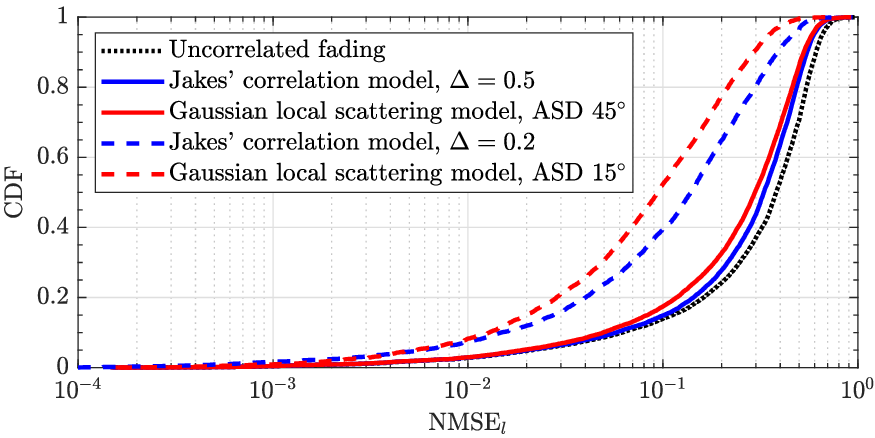}
    \caption{CDF of the per-AP NMSE for different levels of spatial correlation. Settings: $K=20$, $L=64$, and ULA-based APs.}
    \label{fig0102}
\end{figure}

Fig.~\ref{fig0102} shows the CDF of the per-AP NMSE for different channel-correlation models. The Jakes model, with normalized antenna spacings $\Delta = \{0.2, 0.5\}$, is compared with the Gaussian local scattering model, assuming two different sets of angular standard deviations (ASD) for both azimuth and elevation angles: $15^\circ$ and $45^\circ$, which represent the cases for strong and mild correlation, respectively. The figure also shows the reference case of uncorrelated Rayleigh fading with $\bR_{k,l} = \bI_{NQ}$. 
As shown in Fig.~\ref{fig0102}, stronger spatial correlations result in a reduced NMSE, as correlation inherently improves the MMSE estimation quality. 
On the other hand, the uncorrelated Rayleigh fading case represents a lower bound on the performance. The NMSE resulting from the Jakes model is comparable to that obtained by the Gaussian local scattering model with an ASD of 45\textdegree, assuming equal antenna spacing ($\Delta = 0.5$). This result indicates that the Jakes model, when applied to antenna ports spaced half a wavelength apart, results in low correlated fading. The reasoning behind this can be traced to the fundamental physical assumptions of Jakes' model: the signal arrives with uniform angular distribution in azimuth and elevation (isotropic scattering); the phases of multipath components are uniformly random. As a result, the correlation depends only on the distance and not on array orientation or angular spread. So Jakes’ model implicitly assumes all AP antennas see identical, isotropic scattering, which underestimates spatial correlation.

\begin{figure}[t]	
    \centering
    \includegraphics[width=.95\columnwidth]{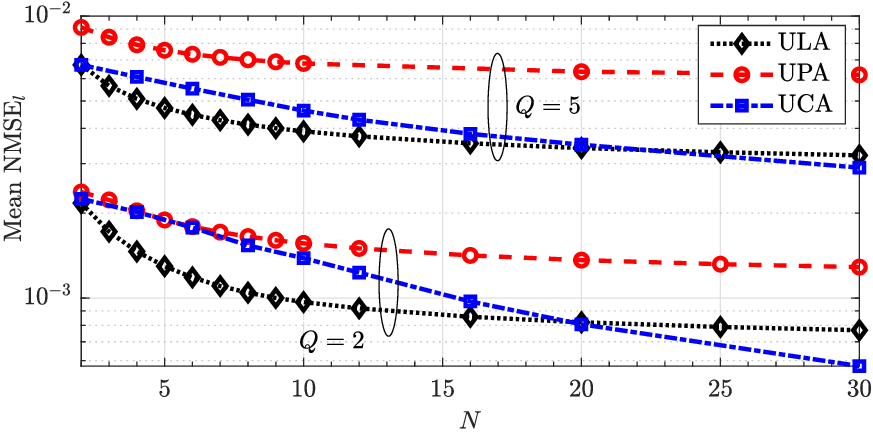}
    \caption{Mean NMSE versus number of antennas per AP. Settings: $K=5$, $L=64$, and $\Delta=0.2$.}
    \label{fig06}
\end{figure}
Fig.~\ref{fig06} shows the mean NMSE$_l$, averaged over several network snapshots, versus the number of FAs per AP for different array geometries. Results are presented for two values of the number of ports per FA, $Q= \{2, 5\}$. The FA arrangement affects the spatial correlation, and hence the channel estimation quality.
The figure indicates that the NMSE achieved by ULA-equipped APs is consistently lower than that of UPA-equipped APs, and also lower than that of UCA-equipped APs for small $N$. While the NMSE achieved by ULA-equipped APs eventually saturates as $N$ increases, the NMSE achieved by UCA-equipped APs significantly reduces. UCA-equipped APs provide a higher channel estimate quality for large $N$, i.e., for more than 20 FAs per AP.
When increasing $N$, while keeping $Q$ fixed, the antenna-port spacing remains constant in ULA and UPA geometries, whereas in UCA, the ports become closer together, increasing spatial correlation and improving estimation quality.
Increasing the number of ports, while maintaining its size, lengthens every single antenna and increases the number of unknowns that need to be estimated. Therefore, the number of ports per FA, $Q$, and the NMSE grow in tandem.

\begin{figure}[t]	
    \centering
    \includegraphics[width=.95\columnwidth]{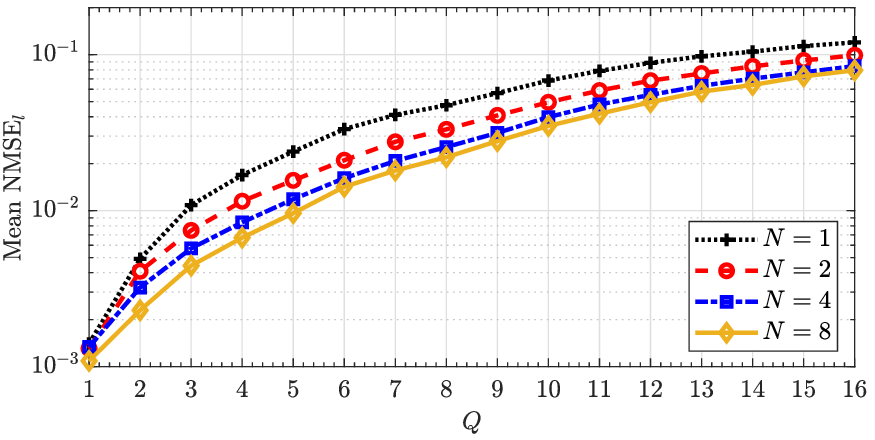}
    \caption{Mean NMSE versus number of ports per FA. Settings: $K=5$, $L=64$, $\Delta=0.2$, Jakes' model and ULA-based APs.}
    \label{fig08}
\end{figure}
This behavior is more rigorously shown in Fig.~\ref{fig08}, which reports the mean NMSE$_l$ against $Q$ for different values of antennas per AP, $N = \{1,\,2,\,4,\,8\}$. In these simulations, we consider Jakes' correlation model and ULA-based APs. While the NMSE increases with $Q$, as motivated above, it decreases mildly with $N$. This improvement occurs because the channel realizations at adjacent (and closer) antennas are highly correlated, which the MMSE estimator leverages to enhance estimation quality. However, the gains from increasing $N$ are minimal for larger values because the ULA becomes so wide that the channel realizations at the outermost antennas are nearly uncorrelated.

\begin{figure}[t]	
    \centering
    \includegraphics[width=.95\columnwidth]{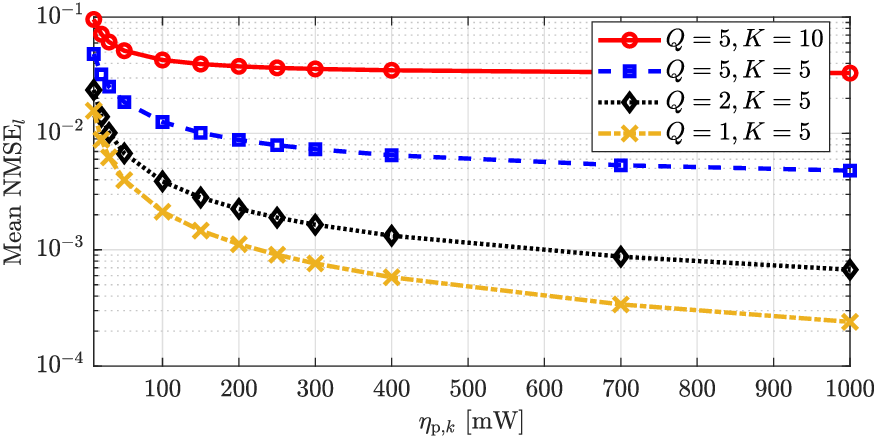}
    \caption{Mean NMSE versus user transmit power. Settings: $L=64$, $\Delta=0.2$, Jakes' correlation model and ULA-based APs.}
    \label{fig02}
\end{figure}
Fig.~\ref{fig02} illustrates the mean NMSE$_l$ versus the pilot transmit power per user, $\eta_{\text{p},k}$, in mW, for different numbers of users and antenna ports. As observed, increasing the number of users or antenna ports increases the mean NMSE due to increased multi-user interference in the pilot observations. Moreover, for $K = 10$, the NMSE curves saturate at high transmit power levels, indicating that beyond a certain threshold, further increases in power do not significantly improve channel estimation accuracy, as this is counterbalanced by increased interference.

\begin{figure}[t]	
    \centering
    \includegraphics[width=.95\columnwidth]{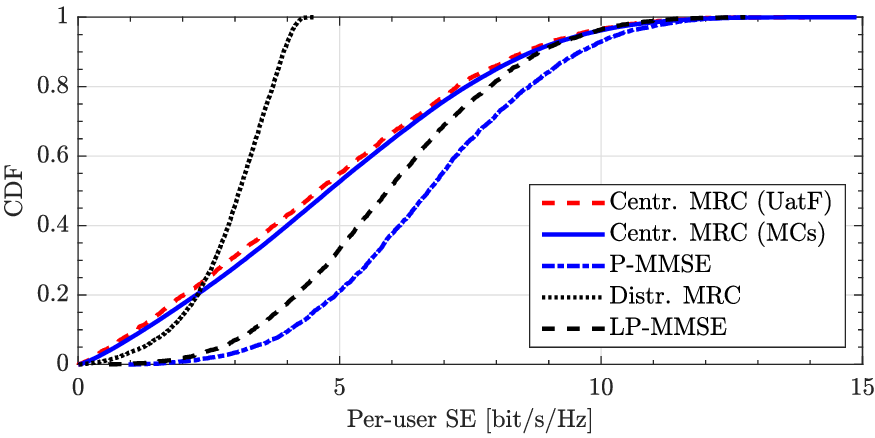}
    \caption{CDF of the per-user SE in different uplink operations. Settings: $L=100$, $K=20$, $\tau_p =20$, full power transmission at each user, Jakes' correlation model, and ULA-based APs.}
    \label{fig04}
    \vspace{-3mm}
\end{figure}

Fig.~\ref{fig04} compares the per-user uplink SE achieved under centralized and distributed operations, as given by \eqref{eq:SE:centralized} and \eqref{eq:SE:distributed}, respectively, using MR and MMSE combiners. The results demonstrate that the choice of combining method has a significant impact on performance. For centralized MRC, the SE attained by using the UatF capacity-bounding technique in~\eqref{eq:ClosedFormUatF}, closely resembles the SE obtained from \eqref{eq:SE:centralized} using Monte Carlo simulations (MCs), as shown in Fig.~\ref{fig04}. In addition, we observe that P-MMSE and LP-MMSE significantly outperform MRC in their respective operation modes, emphasizing the crucial relevance of suppressing multi-user interference in such an interference-limited scenario. As expected, centralized P-MMSE achieves uniformly superior performance compared to distributed LP-MMSE.

\begin{figure}  
    \centering  
    \subfloat[Centralized MRC]{\includegraphics[width=.95\columnwidth]{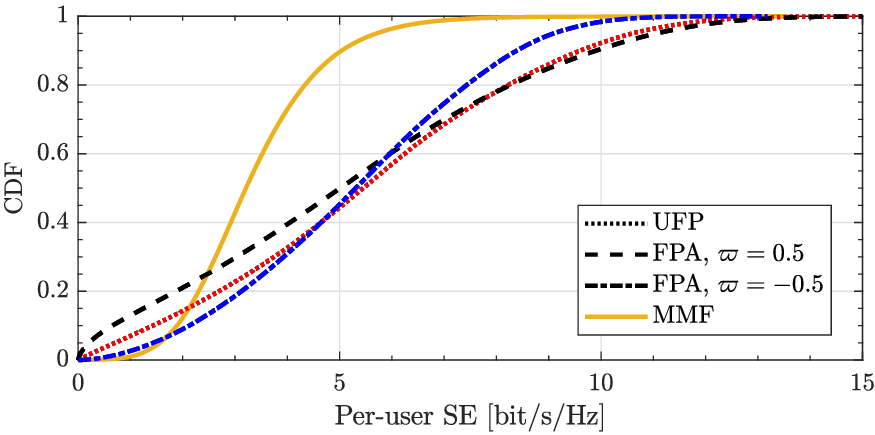} \label{Centralized-MRC}}\\%
    \subfloat[LP-MMSE with n-opt LSFD]{\includegraphics[width=.95\columnwidth]{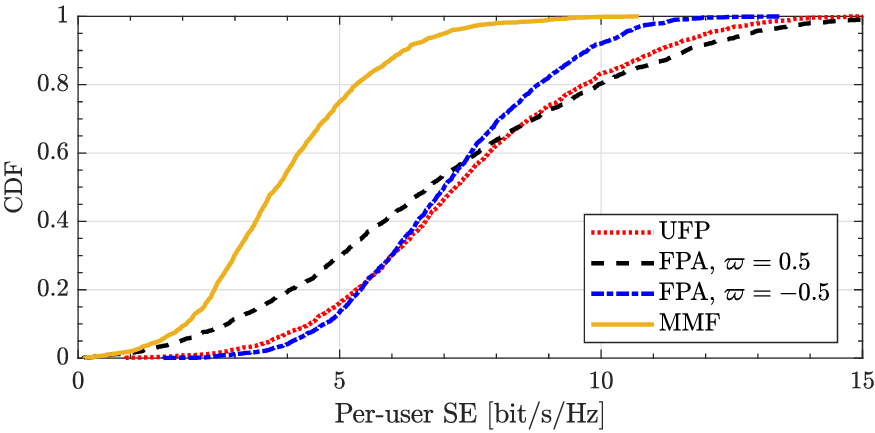} \label{LP-MMSE-LSFD}}%
    \caption{CDF of the per-user SE in the centralized and distributed
uplink operation with different power allocation schemes. Settings: $K=10$, $L=64$, Jakes' correlation model and ULA-based APs.}
    \label{fig05}
\end{figure}

Fig.~\ref{Centralized-MRC} and Fig.~\ref{LP-MMSE-LSFD} illustrate the SE achieved by the centralized MRC and distributed LP-MMSE schemes, respectively, for different user power allocation schemes. In these simulations, we consider the optimal MMF power allocation~\cite[Algorithm 7.1]{demir2021foundations}, the FPA with $\varpi= \{0.5,-0.5\}$, and the UFP (i.e., the FPA with $\varpi=0$). Let us recall that negative values of $\varpi$ imply allocating more power to the users experiencing poor channel gains, thereby implementing an egalitarian power allocation strategy, whereas positive values of $\varpi$ favor users with larger channel gains in an opportunistic fashion. These trends are confirmed by the CDFs of the per-user SE shown in Fig.~\ref{fig05}. Notably, FPA with $\varpi= -0.5$ represents an excellent trade-off between sum SE and user fairness. Indeed, by properly tuning $\varpi$ in the FPA scheme, it can be achieved a 95\%likely per-user SE either comparable (MRC case) or much larger (LP-MMSE case) than that of the optimal MMF power allocation. 

\begin{figure}[t]
    \centering  
    \subfloat[Centralized Operation]{\includegraphics[width=.95\columnwidth]{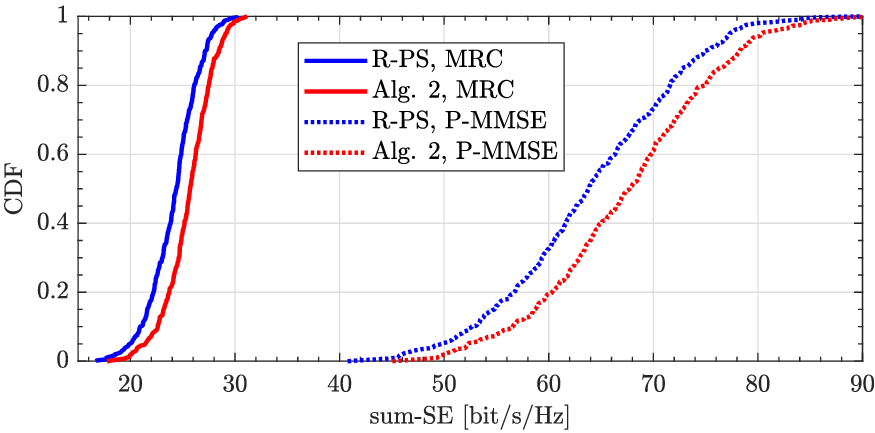} \label{Centralized-Operation}}\\%
    \subfloat[Distributed Operation]{\includegraphics[width=.95\columnwidth]{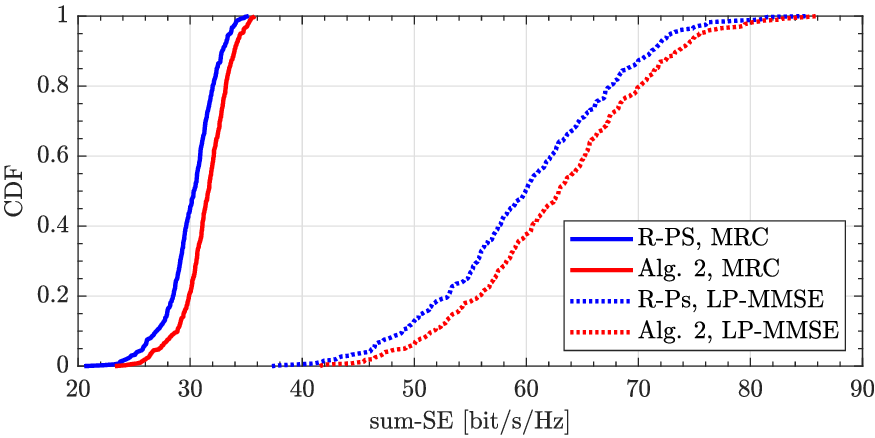} \label{Distributed-Operation}}%
    \caption{CDF of the uplink sum SE for fixed and optimum position. Settings: $K=10$ and $L=64$.}
    \label{fig07}
\end{figure}

Finally, Fig.~\ref{fig07} shows the benefits of optimizing the antenna port positions by using Algorithm~\ref{alg:antenna_position} for both centralized (Fig.~\ref{Centralized-Operation}) and distributed (Fig.~\ref{Distributed-Operation}) operations. The MRC, LP-MMSE, and P-MMSE schemes are compared, assuming FPA with $\varpi =- 0.5$ for all setups.  
Optimizing the antenna-port positions provides significant sum-SE gains over a random port selection (R-PS). Notably, the largest sum-SE improvements are observed for P-MMSE and LP-MMSE, as optimal port selection effectively increases the degrees of freedom to enhance the multi-user interference suppression.
 
\section{Conclusion}
\label{concl}
This study showed that integrating FAs into CF-mMIMO networks significantly improves spatial adaptability and system performance. Optimizing FA port positions increases overall spectral efficiency. We proposed a channel estimation framework with optimized active port selection to minimize the normalized mean squared error. We also analyzed how channel spatial correlation and different AP array geometries affect performance, offering guidelines for practical FA deployment. Future work includes learning-based methods for real-time FA position adaptation and low-complexity distributed estimation schemes to fully exploit FAs.

\appendix
\subsection{Proof of~\eqref{eq:ClosedFormUatF}}

The result follows by evaluating the expectations in~\eqref{UatF}, 
using the MMSE statistics in~\eqref{Ckl}, the combining vector in~\eqref{MR}, 
and the fact that
\begin{equation}
\boldsymbol{\Gamma}_{k,l} 
\triangleq 
\E\{\hat{\mathbf{h}}_{k,l} \mathbf{h}_{k,l}\herm\}
= \E\{\hat{\mathbf{h}}_{k,l} \hat{\mathbf{h}}_{k,l}\herm \}
= \mathbf{R}_{k,l} - \mathbf{C}_{k,l}\label{Gammakl}\,,
\end{equation}
due to the uncorrelation between MMSE estimate and estimation error.
As for the numerator of the SINR in~\eqref{UatF}, assuming MRC combining $\mathbf{v}_{k,l} = \mathbf{P}_{\text{d},l} \hat{\mathbf{h}}_{k,l}$, and using the cyclic property of trace, linearity of expectation, and the MMSE estimation property, we compute
\begin{align}
\EX{\mathbf{v}_{k,l}\herm \mathbf{P}_{\text{d},l} \mathbf{h}_{k,l}}
&= \EX{\hat{\mathbf{h}}_{k,l}\herm \mathbf{P}\herm_{\text{d},l}\mathbf{P}_{\text{d},l} \mathbf{h}_{k,l}} \nonumber \\
&= \tr \big\{ \mathbf{P}_{\text{d},l} \E\{\hat{\mathbf{h}}_{k,l} \mathbf{h}_{k,l}\herm\} \mathbf{P}\herm_{\text{d},l}\big\} \nonumber \\
&= \tr \big\{ \mathbf{P}_{\text{d},l} \boldsymbol{\Gamma}_{k,l} \mathbf{P}\herm_{\text{d},l}\big\}\,\,
\end{align}
where $\boldsymbol{\Gamma}_{k,l}$ is defined in~\eqref{Gammakl2}.
Hence, the numerator of the SINR in~\eqref{UatF} is, closed form, given by
\begin{align}
\textsf{Num}_k = {{\eta}_{\text{d},k}} \left| \sum\nolimits_{l\in \mathcal{L}_k} \tr\{\mathbf{P}_{\text{d},l} \boldsymbol{\Gamma}_{k,l} \mathbf{P}\herm_{\text{d},l}\} \right|^2. 
\label{num}
\end{align}
Similarly, we compute the third term of the denominator in~\eqref{UatF} in closed form as 
\begin{align}
\textsf{No}_k &= \sigma^2 \sum\nolimits_{l \in \mathcal{L}_k} \EX{\|\mathbf{v}_{k,l}\|^2} = \sigma^2 \sum\nolimits_{l \in \mathcal{L}_k} \EX{\mathbf{v}_{k,l}\herm \mathbf{v}_{k,l}} \nonumber \\
&= \sigma^2 \sum\nolimits_{l\in \mathcal{L}_k} \tr\big\{\mathbf{P}_{\text{d},l} \boldsymbol{\Gamma}_{k,l} \mathbf{P}\herm_{\text{d},l}\big\}. 
\label{noise}
\end{align}
Let us focus on the first expectation of the denominator in~\eqref{UatF} 
\begin{align} &\EX { \Big| \sum\nolimits_{l\in \mathcal{L}_k} \mathbf{v}_{k,l}\herm \mathbf{P}_{\text{d},l} \mathbf{h}_{j,l} \Big|^2 } \nonumber \\ 
&\quad=\sum_{l \in \mathcal{L}_k} \sum_{l' \in \mathcal{L}_k} \EX {\hat{\mathbf{h}}_{k,l}\herm \mathbf{P}\herm_{\text{d},l}\mathbf{P}_{\text{d},l} \mathbf{h}_{j,l} \mathbf{h}_{j,l'}\herm \mathbf{P}\herm_{\text{d},l'} \mathbf{P}_{\text{d},l'}\hat{\mathbf{h}}_{k,l'}} \nonumber \\
&\quad=\sum\nolimits_{l\in\mathcal{L}_k} \EX{\hat{\mathbf{h}}_{k,l}\herm {\mathbf{P}\herm_{\text{d},l}} \mathbf{P}_{\text{d},l} \mathbf{h}_{j,l} \mathbf{h}_{j,l}\herm {\mathbf{P}\herm_{\text{d},l}}\mathbf{P}_{\text{d},l} \hat{\mathbf{h}}_{k,l}}\,,
\label{interference-closed-form}
\end{align}
where, in the last equality, the cross-product terms for $l\neq l'$ vanish due to the statistical independence of the channels across different APs.

By using $\E\{\mathbf{x}\herm \mathbf{A} \mathbf{x}\} = \tr\{\mathbf{A} \E\{\mathbf{x}\mathbf{x}\herm\}\}$,~\eqref{interference-closed-form} equals
\begin{align}
&\textsf{Intf}_{k,j} = \sum\nolimits_{l\in\mathcal{L}_k} \tr \Big\{{\mathbf{P}\herm_{\text{d},l}} \mathbf{P}_{\text{d},l} \E\{\mathbf{h}_{j,l} \mathbf{h}_{j,l}\herm {\mathbf{P}\herm_{\text{d},l}}\mathbf{P}_{\text{d},l} \hat{\mathbf{h}}_{k,l} \hat{\mathbf{h}}_{k,l}\herm\} \Big\} \nonumber \\
&\quad = \sum\nolimits_{l\in\mathcal{L}_k} \tr\Big\{{\mathbf{P}\herm_{\text{d},l}}\mathbf{P}_{\text{d},l} \mathbf{R}_{j,l} {\mathbf{P}\herm_{\text{d},l}} \mathbf{P}_{\text{d},l} \boldsymbol{\Gamma}_{k,l}\Big\}\,,
\end{align}
where, in the last equality, we utilized the independence of $\hat{\mathbf{h}}_{k,l}$ and $\mathbf{h}_{j,l}$, $k \neq j$, under the assumption of $\tau_p \geq K$ and no pilot reuse.
Hence, the denominator of the SINR in~\eqref{UatF} of SINR is given by
\begin{align}
\textsf{Den}_k = \sum\nolimits_{j=1}^{K} \eta_{\text{d},j} \textsf{Intf}_{k,j} - \textsf{Num}_k + \textsf{No}_k\,.
\label{interference}
\end{align}
By inserting the closed-form expressions~\eqref{num},~\eqref{noise} and~\eqref{interference} into~\eqref{UatF}, we finally derive~\eqref{eq:ClosedFormUatF}.

\balance

\bibliographystyle{IEEEtran}
\bibliography{references}

\end{document}